\documentclass[11pt]{article}

\usepackage{lineno}
\usepackage[margin=3cm]{geometry}
\usepackage{float}
\usepackage{graphicx}
\usepackage{authblk}
\usepackage{caption}
\usepackage{subcaption}
\usepackage{longtable}
\usepackage{xcolor}
\usepackage{wrapfig}
\usepackage{tabularx}
\usepackage{authblk}
\usepackage[title, titletoc]{appendix}
 
\usepackage{amssymb,amsmath,amsthm}

\newtheorem{assumption}{Assumption}
\newtheorem{definition}{Definition}
\newtheorem{theorem}{Theorem}
\newtheorem{remark}{Remark}

\usepackage{booktabs}
 \allowdisplaybreaks
\usepackage{stmaryrd}
\usepackage{url}

\usepackage{endnotes}
\let\footnote=\endnote

\usepackage{booktabs, longtable, enumitem, mathtools, ulem}
\usepackage[mode=buildnew]{standalone}
\usepackage[colorlinks=true,breaklinks=true,bookmarks=true,urlcolor=blue,
     citecolor=blue,linkcolor=blue,bookmarksopen=false,draft=false]{hyperref}
     
\newcommand{\B}{\mathcal{B}}
\newcommand{\I}{\mathcal{I}}
\newcommand{\J}{\mathcal{J}}
\renewcommand{\L}{\mathcal{L}}
\newcommand{\W}{{\bf W}}
\renewcommand{\r}{{\bf r}}
\newcommand{\s}{{\bf s}}
\newcommand{\w}{{\bf w}}
\newcommand{\st}{\operatorname{st.}}
\newcommand{\R}{\mathbb{R}}
\newcommand{\dsum}{\displaystyle \sum}
\renewcommand{\dfrac}{\displaystyle \frac}
\newcommand{\ccozeta}{\zeta}

\newcommand{\sched}{sch}
\newcommand{\spill}{spi}
\newcommand{\argmin}{\operatorname{argmin}}

\newcommand{\case}[1]{{\tt Case:#1}}     

\usepackage{natbib}
 \bibpunct[, ]{(}{)}{,}{a}{}{,}%

\title{Revenue Adequate Prices for Chance-Constrained Electricity Markets with Variable Renewable Energy Sources}

\author[2]{Xin Shi\thanks{\texttt{xis316@lehigh.edu}}}
\author[1,2]{Alberto J. Lamadrid L.\thanks{Corresponding Author:\texttt{ajlamadrid@ieee.org}}}
\author[2]{Luis F. Zuluaga\thanks{\texttt{luis.zuluaga@lehigh.edu}}}
\affil[1]{Department of Economics, Lehigh University, Bethlehem, PA, USA}
\affil[2]{Department of Industrial and Systems Engineering, Lehigh University, Bethlehem, PA, USA}

\date{\vspace{-5ex}}

\begin{document}

\maketitle

\begin{abstract}
In a commodity market, revenue adequate prices refer to compensations that ensure that a market participant has a non-negative profit. In this article, we study the problem of deriving revenue adequate prices for an electricity market-clearing model with uncertainties resulting from the use of variable renewable energy sources (VRES). To handle the uncertain nature of the problem, we use a chance-constrained optimization (CCO) approach, which has recently become very popular choice when constructing dispatch electricity models with penetration of VRES (or other sources of uncertainty).  Then, we show how prices that satisfy revenue adequacy in expectation for the market administrator, and cost recovery in expectation for all conventional and VRES generators, can be obtained from the optimal dual variables associated with the deterministic equivalent of the CCO market-clearing model. These results constitute a novel contribution to the research of research on revenue adequate, equilibrium, and other types of pricing schemes that have been derived in the literature when the market uncertainties are modeled using stochastic or robust optimization approaches. Unlike in the stochastic approach, the CCO market-clearing model studied here produces uncertainty uniform real-time  prices that do not depend on the real-time realization of the VRES generation outcomes. 
To illustrate our results, we consider a case study electricity market, and contrast the market prices obtained using a revenue adequate stochastic approach and the proposed revenue adequate CCO approach.
\end{abstract}


\section{Introduction}

In a commodity market, market-clearing prices refer to prices that allow the commodity producers and consumers (or buyers and sellers) to trade the commodity so no surplus or deficit occurs. Classical economic theory studies exchanges of goods and services, with common terms for all participating agents seeking equilibrium between supply and demand. Sometimes these common terms take the form of centralized auctions. Centralized auction mechanisms have been used for allocation in a wide range of areas. This includes envisioned electricity markets before their actual implementation~\citep{schweppe1988}, electromagnetic spectrum license allocations for the US Federal Communications Commission \citep{milgrom2000, fox2013}; emissions of sulfur dioxide under Title IV of the 1990 Clean Air Act Amendments \citep{joskow1998}; and power exchanges in wholesale electricity markets \citep{hortacsu2008}.

Obtaining appropriate market-clearing prices for electricity markets is a challenging problem due the way in which electricity is produced, transmitted and consumed, as well as the fact that electricity, loosely speaking, cannot yet be economically stored. Thus, in many electricity markets (e.g., California ISO, PJM Interconnection, New York ISO, New England ISO, ERCOT, and Nordpool), a market administrator is tasked with the central administration of the market that is referred as the independent system operator (ISO).
One of these challenges arises from the increasing penetration of variable renewable energy sources (VRES); like solar, wave, and wind generators, in electricity markets, following efforts to move towards a low carbon economy. Namely, the uncertain and intermittent nature of the power generated by the VRES generating units introduces uncertainty in the market-clearing model used to compute the desired market-clearing prices. 

\looseness=-1
Typically, market-clearing models for electricity markets are formulated as two-stage models~\citep[see, e.g.,][]{khazaei2013market, bjorndal2016congestion} with matching settlements. In the market's scheduling (first) stage, usually timed a day-ahead before the market's real-time (second) stage, conventional and VRES generators make offers, and the market administrator chooses scheduled (or pre-dispatch) quantities of electricity generation. 
Then, in the market's real-time stage, when delivery of power occurs, new sets of generation bids are submitted (e.g., by VRES generators), which  can
deviate (i.e., be redispatched) from the scheduled dispatch levels to clear the market. As considered here, in the real-time stage, the market administrator might also curtail the market loads (demands), subject to compensation to consumers. The market administrator's objective in setting the scheduled dispatches, and the real-time dispatches and curtailments, is to maximize the social welfare. 

\looseness=-1
In deterministic market-clearing models in the literature, the uncertain values (e.g., VRES generation) in the scheduling stage are replaced by their expected values, and both the scheduling and the real-time stage social welfare maximization problems are run separately in a two-settlement fashion~\citep{khazaei2013market}. These models can be improved by better modeling the uncertainties in the electricity market and co-optimizing the scheduling and real-time stages~\citep{bjorndal2016congestion}. 

The most popular way to do this, is by developing stochastic market-clearing models. That is, models in which the market uncertainties are modeled by considering their distribution to be given by a finite set of  scenarios with corresponding probabilities of occurrence. This choice leads stochastic market-clearing models to be formulated as two-stage stochastic optimization problems. In particular, consider the work of~\citet{wong2007pricing, pritchard2010single, morales2012pricing, zavala2017stochastic, kazempour2018stochastic, bose2015design}, who develop {\em revenue adequate} market-clearing prices (discussed next) in expectation; \citet{abbaspourtorbati2016pricing, zakeri2018pricing}, who develop revenue adequate market-clearing prices in each scenario; and \citet{khazaei2017single} who develop equilibrium market-clearing prices; to name just a few. 

The use of stochastic optimization techniques~\citep[see, e.g.,][]{birge2011introduction} to obtain better market-clearing prices in the presence of market uncertainties is not surprising given the number of stochastic optimization models related, in general, to the dispatch of electricity that have been developed in the literature. For example, consider the work of \citet{pereira1991multi, bouffard2005market, sen2006stochastic,  lamadrid2015, zou2018partially}, to name just a few ones, as well as the reviews given in~\citet{zhou2016stochastic} and \citet{lamadrid2019}. However, uncertainty in electricity models has also been addressed by using {\em robust}~\citep[see, e.g.,][]{ben2009robust} and {\em chance-constrained}~\citep[see, e.g.,][]{nemirovski2006convex} optimization techniques. Some examples of robust optimization models for problems in electricity are the work of  \citet{roald2015security, ding2016robust, lorca2016multistage, chen2018distributionally}. However, to the best of our knowledge, only \citet{kramer2018strictly, ye2016uncertainty} consider the problem of finding equilibrium market-clearing prices for robust market-clearing models. 

The case is similar for chance-constrained optimization dispatch models that have popularized lately in the literature. For example consider the work of  \citet{ozturk2004solution, pozo2012chance, bienstock2014chance, lubin2015robust,zhang2017chance, xu2018optimal,kuang2018pricing, halilbavsic2018convex, lubin2019chance,
venzke2020chance, roald2016corrective, jabr2013adjustable}, to name just a few. However, to the best of our knowledge, only a couple of articles address the problem of designing market-clearing prices for uncertain electricity markets in which the market's uncertainties are handled using chance-constrained optimization. In particular, \citet{wang2011wind, mazadi2013impact, kuang2018pricing}, provide {\em equilibrium} market-clearing prices~\citep[see, e.g.,][]{scarf1990mathematical, ref.oneill2005}. Also, \citet{dvorkin2019chance} extends the results in~\citet{kuang2018pricing} to the case in which convex quadratic cost are incurred in the market, and shows that if a particular (not controllable) dual value of the market's optimization model turns out to be non-negative, then the prices are revenue adequate for the market administrator (see discussion after~\citep[][eq. (23a)]{dvorkin2019chance}). As discussed in more detail next (Section~\ref{sec:contributions}), our main contribution is to develop revenue adequate market-clearing prices for uncertain electricity markets in which the market's uncertainties are handled using chance-constrained optimization.

In the market-clearing prices literature, revenue adequacy refers to prices that ensure that the profit of the market administrator is non-negative; that is, the market administrator does not incur any losses by operating the market. Related to revenue adequacy is the concept of cost recovery. Prices that satisfy cost recovery are those that ensure that the revenue of a market participant is enough to recover its costs~\citep[see, e.g.,][]{morales2012pricing}. 
At times, here we simply refer to prices that satisfy revenue adequacy and cost recovery as revenue adequate prices. The main advantages of revenue adequate prices against more classical {\em equilibrium} prices~\citep[see, e.g.,][]{scarf1990mathematical, ref.oneill2005, kuang2018pricing} are that {\em uplift} payments~\citep[see, e.g.,][]{jayantilal2001market} are not needed to ensure that generators will never be forced to lose money; and that the market administrator is guaranteed to have non-negative operating profits. As mentioned above, the problem of obtaining revenue adequate prices for stochastic market-clearing models has and continues to be widely studied in the literature.

\subsection{Contributions}
\label{sec:contributions}

Our main contribution is to develop a chance-constrained market-clearing model and corresponding chance-constrained pricing scheme that ensures revenue adequacy for the market administrator in expectation, and cost recovery in expectation for all the conventional and VRES generators in the market. To our knowledge, this is the first time that a pricing scheme of this type is developed. Thus,
this pricing scheme paves the way for the design of market-clearing pricing schemes with different properties, for different electricity market-clearing models in which it is appropriate or desirable to model the market's uncertainties through the use of chance-constrained optimization. As mentioned earlier, such chance-constrained market-clearing models are widely considered and studied in recent literature~\citep[see, e.g.][]{ozturk2004solution, pozo2012chance, jabr2013adjustable, bienstock2014chance, lubin2015robust,roald2016corrective, zhang2017chance, xu2018optimal,kuang2018pricing, halilbavsic2018convex, lubin2019chance,
venzke2020chance}.

Also, by analyzing the market-prices obtained via stochastic and chance-constrained market-clearing models, we show that there is a fundamental difference between these two pricing schemes. Namely, stochastic pricing schemes derive prices for the real-time participants' actions that might change depending on the realization of the market's uncertain parameter(s) (e.g., VRES generation). In contrast, the proposed chance-constrained market-clearing model and associated market-clearing pricing scheme result in uncertainty uniform prices for the real-time market participants' actions that do not change depending on the realization of the market's uncertain parameter(s). This feature circumvents governance difficulties when implementing a stochastic market-clearing pricing scheme in electricity markets; for example, having all stakeholders to agree on a set of scenarios to use and their associated welfare compensations. 
Moreover, we show that the use of the dual variables associated with the market-clearing model beyond those related to the scheduling and real-time stage balancing constraints, allow to obtain distributions of the revenues throughout the network that can be significantly different to the ones resulting from a stochastic market-clearing pricing scheme. 
The rest of the article is organized as follows. In Section~\ref{sec:prelim}, we present most of the notation to be used throughout the article. Also, we introduce a deterministic market-clearing model for a market administrator of an electricity market with participating conventional generators, VRES generators, and demand side representatives (e.g., load serving entities). This model is equivalent to the deterministic or nominal version of the well-known stochastic market-clearing model studied in~\citet{morales2012pricing}. This fact helps us to contrast the latter model with the chance-constrained market-clearing model studied here. In Section~\ref{sec:CCmodel}, we present a chance-constrained market-clearing model associated with the deterministic model in Section~\ref{sec:prelim}, when one takes into account the uncertain nature of VRES generation. In Section~\ref{sec:CCPricing}, we present a pricing scheme for this chance-constrained market-clearing model that ensures revenue adequacy for the market administrator, and cost recovery for the conventional and VRES generators participating in the market. In Section~\ref{sec:casestudy}, we present numerical results for an electricity market case study. In Section~\ref{sec:vsstoc}, we contrast some of the properties of revenue adequate prices obtained via a stochastic vs a chance-constrained market-clearing pricing scheme. In Section~\ref{sec:final}, we provide some final remarks. For the purpose of brevity, we have included important, but less relevant information regarding model assumptions, and additional information about the stochastic market-clearing scheme in~\citet{morales2012pricing}, to appendices referenced throughout the article. 

\section{Preliminaries: Notation and Nominal Model}
\label{sec:prelim}
In this section, we begin by introducing most of the notation used throughout the article. Then, we introduce the nominal (i.e., deterministic) version of the electricity market-clearing model. This deterministic market-clearing model is used to develop the uncertain version of the market-clearing model based on a chance-constrained (CC) optimization (CCO) approach. For this uncertain market clearing-model, we derive a pricing scheme satisfying revenue adequacy (or cost recovery) in expectation for all power producers and the market administrator. The nominal model is equivalent to the nominal model behind the two-stage stochastic optimization (SO) market-clearing model considered by~\citet{morales2012pricing}. This fact allows us to contrast the properties of the SO pricing scheme and the CCO pricing scheme derived here. 
Tables~\ref{tab:setnomenclature},~\ref{tab:parnomenclature}, and~\ref{tab:varnomenclature} present the sets, parameters, and decision variables used to introduce the nominal (i.e., deterministic) version of the electricity market model considered here. 
\begin{table}[!htb]
\caption{Sets associated with the market-clearing optimization model.}
\label{tab:setnomenclature}
\footnotesize
\begin{center}
\begin{tabular}{lp{0.85\linewidth}}
\toprule
\multicolumn{2}{c}{\bf Sets}\\
\midrule
    $\B$ & {Set of buses}\\
			$\I$ & {Set of conventional generators}\\
			$\I_n$ & {Set of conventional generators $\I_n \subseteq I$ at bus $n \in \B$}\\
			$\J$ &{Set of loads}\\
			$\J_n$ & {Set of loads $\J_n \subseteq \J$ at bus  $n \in \B$}\\
$\L $ & {Set of transmission lines $\L \subseteq \{(i,j) \in \B \times \B\}$} (satisfying $(i,j) \in \L \Rightarrow (j,i) \in \L$)\\
\bottomrule
\end{tabular}
\end{center}
\end{table}

\begin{table}[!htb]
\caption{Parameters associated with the market-clearing optimization model.}
\label{tab:parnomenclature}
\footnotesize
\begin{center}
\begin{tabular}{lp{0.85\linewidth}}		    
		    \toprule
		    \multicolumn{2}{c}{\bf Parameters}\\
\midrule
		   	 $b(k)$ & {Bus $b(k) \in \mathcal{B}$ where generator $k \in \mathcal{I}$, or load $k \in \J$, is located}\\
			 $B_{kl}$ & {Absolute value of the susceptance of line $(k,l) \in \L$}\\
			 $C_i$ & {Per unit generation cost of generator $i \in \I $}\\
			 $C_i^u$ & {Per unit cost for upward reserve deployment of generator $i \in \I$}\\
			 $C_i^d$ & {Per unit saving for downward reserve deployment of generator  $i \in \I$}\\
			 $C_n^w$ & {Per unit cost of VRES power generation at bus $n \in \B$}\\
			 $\overline{C}_{kl}$ & {Transmission capacity of line $(k,l) \in \L$}\\
			 $L_j$ & {Scheduled power consumption by load $j \in \J$}\\
			 $\overline{P}_i$ & {Maximum generation capacity of conventional generator $i \in \I$}\\
			 $\overline{R}_i^u$ & {Maximum upward reserve that can be provided by conventional generator  $i \in \I$}\\
			 $\overline{R}_i^d$ & {Maximum downward reserve that can be provided by conventional generator $i \in \I$}\\
			 $V_j$  & {Per unit cost of involuntary curtailment of scheduled load $j \in \J$}\\
			 $W_n$ & {VRES power generation at bus $n \in \B$}\\
		     $\overline{W}_n$& {Maximum allowed scheduled generation by VRES generator $n \in \B$}\\
		     \bottomrule
		     \end{tabular}
		     \end{center}
		     \end{table}	
		     
\begin{table}[!htb]
\caption{Variables associated with the market-clearing optimization model.}
\label{tab:varnomenclature}
\footnotesize
\begin{center}
\begin{tabular}{lp{0.85\linewidth}}				     
		     \toprule
\multicolumn{2}{c}{\bf Variables}\\
\midrule	
 $\delta_{n}^0$&{Scheduled voltage angle at bus $n \in \B$}\\
		     $\delta_{n}$ & {Real-time voltage angle at bus $n \in \B$}\\
			 $p_i $& {Scheduled power generation of conventional generator $i \in \I$}\\
			 $r_i^u$ & {Real-time upward reserve deployed by generator $i \in \I$}\\
			 $r_i^d$ &{Real-time downward reserve deployed by generator $i \in \I$}\\
			 $s_j$ & {Real-time involuntary curtailment of scheduled load $j \in \J$}\\
			 $w_n^{\sched}$ & {Scheduled VRES power generation at bus $n \in \B$}\\
			 $w_n^{\spill}$ & {Real-time VRES power spilled at bus $n \in \B$}\\
\bottomrule
\end{tabular}
\end{center}
\end{table}
In Table~\ref{tab:setnomenclature}, the condition on the set of lines $\L$ is made for brevity in expressing power balance constraints therein (c.f., ~\eqref{eq:deter_balance},~\eqref{eq:deter_reblanace} below).

We now introduce the nominal (i.e., deterministic) two-stage electricity market-clearing model using a DC approximation of the power flow equations. Namely,
	\begin{subequations}\label{eq:detmodel}
		\begin{align}
		\min &&& \sum_{i \in \I}(C_i p_i + C_i^u r_i^u - C_i^d r_i^d)+ \sum_{n \in \B} C_n^w (W_n - w_n^{\spill}) + \sum_{j \in \J} V_js_j \label{eq:deter_obj} \\
		\st   &&&\sum_{i \in \I_n} p_i + w_n^{\sched}  - \sum_{(n,l) \in \L} B_{nl} (\delta_{n}^0 -  \delta_{l}^0) =\sum_{j \in \J_n}L_j , \label{eq:deter_balance} & \forall n\in \B\\
		&&& \sum_{i \in \I_n}(r_i^u - r_i^d) + \sum_{j \in \J_n} s_j + W_n - w_n^{\sched}- w_n^{\spill}+  \nonumber \\
		&&& \hspace{2in} \sum_{(n,l) \in \L} B_{nl} (\delta_{n}^0 - \delta_{n} - \delta_{l}^0 + \delta_{l}) = 0,   \label{eq:deter_reblanace} & \forall n\in \B  \\
		&&& B_{kl}(\delta^0_{k} - \delta^0_{l}) \leq   \overline{C}_{kl}, \label{eq:deter_line1} & \forall (k,l) \in \L \\ 
		&&& B_{kl}(\delta_{k}-\delta_{l}) \leq   \overline{C}_{kl}, \label{eq:deter_line2} & \forall (k,l) \in \L   \\
		&&&  0 \le w_n^{\sched} \leq  \overline{W}_n, \label{eq:deter_wind_limit} &  \forall n \in \B     \\ 
		&&&  0 \le p_i  \leq \overline{P}_i,\label{eq:deter_con_limit} & \forall i \in \I    \\
		&&&  0 \le w_n^{\spill} \leq  W_n, \label{eq:spill_wind_limit} &  \forall n \in \B     \\ 
		&&& 0 \le r_i^u\leq \overline{R}_i^u ,\label{eq:deter_rampup_limit} & \forall i \in \I    \\
		&&& 0 \le r_i^d \leq  \overline{R}_i^d, \label{eq:deter_rampdown_limit} & \forall i \in \I     \\
		&&& 0 \le p_i +  r_i^u - r_i^d \leq \overline{P}_i, \label{eq:deter_con2_limit} & \forall i \in \I     \\
		&&& 0 \le s_j \leq L_j. \label{eq:deter_demand_limit} & \forall j \in \J    
		\end{align}
	\end{subequations}
In the electricity market-clearing model~\eqref{eq:detmodel}, the objective is to maximize the social welfare of the participants. Specifically, the objective function~\eqref{eq:deter_obj} is to minimize the cost of the power system operation, where: $\sum_{i \in \I}(C_i p_i + C_i^u r_i^u - C_i^d r_i^d)$ is the cost associated with the operation of conventional generators; $\sum_{n \in \B} C_n^w (W_n - w_n^{\spill})$ is the cost associated with the operation of VRES generators; and $\sum_{j \in \J} V_js_j $ is the cost associated to load curtailments. Constraints~\eqref{eq:deter_balance} and~\eqref{eq:deter_reblanace} correspond to the bus power balance constraints
at the scheduling and real-time stages. Constraints~\eqref{eq:deter_line1} and~\eqref{eq:deter_line2} correspond to the transmission capacity constraints at the scheduling and real-time stages. Constraints \eqref{eq:deter_wind_limit} and \eqref{eq:spill_wind_limit}, respectively set lower and upper bounds for the scheduled VRES power generation and the real-time VRES power spilled. To simplify the exposition, throughout we assume that each bus $n \in \B$ has VRES power generation capacity. However, all the results can be adapted in straightforward fashion to the case in which there are buses in the network without VRES power generation capacity (see, Remark~\ref{rem:zero} in Appendix~\ref{sec:assumptions} for details).
Constraints~\eqref{eq:deter_con_limit} set lower and upper bounds for the scheduled conventional power generation. Constraints~\eqref{eq:deter_rampup_limit} and~\eqref{eq:deter_rampdown_limit} set lower bounds and upper bounds on the reserve deployed by each conventional generator. In particular, the upper bounds ensure that these reserves, deployed in the real-time stage, do not exceed their reserve capacity. Constraints~\eqref{eq:deter_con2_limit} ensure that the generation of each conventional generator is not negative and stays within their respective capacities. Constraints \eqref{eq:deter_demand_limit} set lower and upper bounds in the amount by which each scheduled load can be curtailed. For ease of presentation, we have not included in problem~\eqref{eq:detmodel} the reference bus constraints $\delta^0_1=0$, $\delta_1 = 0$, as these constraints do not affect any of the results regarding the pricing schemes discussed in Section~\ref{sec:CCPricing}, and Appendix~\ref{sec:stocmodel}. These constraints are however included when obtaining the numerical results in Section~\ref{sec:casestudy}, and Section~\ref{sec:numerical}. 
 
Model~\eqref{eq:detmodel} corresponds to the nominal (deterministic) optimization model associated with the two-stage stochastic market-clearing model used by~\citet{morales2012pricing} to obtain market prices that ensure revenue adequacy in expectation when there is uncertainty in the market due to the penetration of VRES; for example, wind, solar, and wave energy. Besides the DC approximation, Model~\eqref{eq:detmodel} uses other typical simplifications considered in market-clearing models. For brevity, we use the simplifications discussed in~\citet[][assumptions 1)-6), Sec.~II]{morales2012pricing} and refer the reader to that work. 
On the other hand, as discussed in Section~\ref{sec:CCmodel}, instead of using an atomic distribution of the potential VRES power generation outcomes~\citep[cf.,][assumption 9), Sec.~II]{morales2012pricing},  uncertainty in VRES power generation is added to the deterministic model~\eqref{eq:detmodel} by using distributional information on its forecasting error. This is done without sacrificing the correlation information between different VRES generators (see beginning of Section~\ref{sec:CCmodel}). The way the uncertainty of the VRES generation is modeled here is one of the main differences between the SO approach studied in~\citet{morales2012pricing} and the CCO approach presented here for an uncertain market-clearing model, and leads to the derivation of a pricing scheme with {\em uncertainty uniform} real-time prices that do not depend on the real-time realization of the VRES generation outcomes. 
Similar to~\citet{morales2012pricing, bienstock2014chance, lubin2015robust}, we consider the uncertainty in the electricity market to be associated with the VRES generators. However, the results discussed therein apply similarly when considering uncertain demand, whose uncertainty can be similarly modeled.

\section{Chance-Constrained model of Uncertain Energy Market}
\label{sec:CCmodel}
In this section, the deterministic model~\eqref{eq:detmodel} is used to derive a CCO market-clearing model of an uncertain energy market. The uncertainty is introduced by considering that the (real-time) VRES power generation $\W_n$ at each bus $n \in \B$ cannot be precisely estimated. In particular, similar to~\citet{bienstock2014chance}, we assume that the VRES power generation at each bus $n \in \B$ of the network is a random variable. Namely,
\begin{equation}
\label{eq:winddef}
\W_n = W_n^f + \Delta \W_n^f, \text{with }  \Delta \W_n^f \sim \mathcal{N}(0, \sigma_n > 0),
\end{equation}
that is, with $\Delta \W_n^f $ being 
normal random variables with mean zero  and standard deviation $\sigma_n > 0$  for each~$n \in \B$. In~\eqref{eq:winddef},~$\Delta \W_n^f$ represents the random forecasting error on~$W_n^f$, the VRES power generation forecasted at bus~$n \in \B$.  

The uncertainty modeling in~\eqref{eq:winddef} is similar to the one used in related CC literature like~\citet[][Section II.A]{pozo2012chance}, \citet[][Section 1.4]{bienstock2014chance}, \citet[][Section II.A]{lubin2015robust}). The difference is that in these works, the (assumed to be known) mean VRES power generation is used, instead of the forecasted VRES power generation in $\eqref{eq:winddef}$. We make this choice to take advantage of the fact that $W_n^f$ is clearly known during the scheduling stage, and VRES power forecasting errors have been studied in detail~\citep[see, e.g.,][for the case of wind generation]{lange2005uncertainty, hodge2011wind}. However, all the results discussed therein hold in straightforward fashion if one uses the uncertainty modeling used in~\citet{pozo2012chance, bienstock2014chance,lubin2015robust}, after changing $W_n^f \to \mathbb{E}(\W_n)$, and $\Delta \W_n^f \to \Delta \W_n$, where $\mathbb{E}(\cdot)$ represents the expectation of the random variable $(\cdot)$, and $\Delta \W_n$ stands for the deviations of the VRES power $\W_n$ from its expected value $\mathbb{E}(\W_n)$.
\citet{bienstock2014chance}, and~\citet{dvorkin2015uncertainty} have relevant discussions about the appropriateness of the assumptions discussed above when expressing the VRES power generation uncertainties in terms of its expected value and deviations from it. However, in our setting, most of these assumptions are unnecessary.

\begin{remark}
As we discuss in Appendix~\ref{sec:assumptions}, most of the assumptions (including the normal distribution assumption) made in the modeling of the VRES power generation in~\eqref{eq:winddef} are made for ease of presentation.
\end{remark}

Note that even if the VRES power forecast errors of different VRES generators turn out to be independent, the outcomes of a statistical estimation of the VRES power forecast values $W_n^f$ can be correlated within different buses $n \in \B$; for example, if the meteorological conditions used to obtain the VRES power forecast values in different buses are similar. This means, in reference to~\citet[][assumption 9), Sec.~II]{morales2012pricing}, that correlation information between the power generated by different VRES is not sacrificed in~\eqref{eq:winddef}. 
The uncertainty modeling choice in~\eqref{eq:winddef} leads to the main difference between the market-clearing pricing scheme proposed here and other related SO market-clearing pricing schemes results in the literature. Unlike in the SO~\citep[see, e.g.,][]{morales2012pricing}, the CCO market-clearing model studied here produces uncertainty uniform real-time  prices that do not depend on the real-time realization of the VRES generation outcomes. Also, unlike in typical robust optimization approaches~\citep[see, e.g.,][]{kramer2018strictly, morales2013integrating}, we take advantage of the knowledge of the VRES power generation forecasting error distribution to avoid the market-clearing results to be overly conservative.

Once the random nature of the VRES power generated is added to the problem, and similar to~\citep[][Section 1.3]{bienstock2014chance}, we introduce {\em affine controls}  to model how real-time decisions related to upward and downward reserve deployment, VRES power spill, and curtailment, change in real-time to adapt to uncertainties related to errors in the VRES power generation forecast.  As discussed in~\citep[][among others]{ jabr2013adjustable, bienstock2014chance} these affine controls provide a good representation of the automatic generation
control (AGC) commonly used in system operation. 
Specifically, for all 
$n \in \B$, $i \in \I$, $j \in \J$, let:
\begin{subequations}
\label{eq:controls}
\begin{align}
 \r_i^u & =  r_i^u - \alpha_i^u \Delta \W_{b(i)}^f, \label{eq:controlsrup}\\
 \r_i^d & =  r_i^d + \alpha_i^d \Delta \W_{b(i)}^f, \label{eq:controlsrdo}\\
  \w_{n}^{\spill} & =  w_n^{\spill} + \beta_n \Delta \W_{n}^f, \label{eq:controlssp}\\
 \s_j & =  s_j - \gamma_j \Delta \W_{b(j)}^f, \label{eq:controlscu}
\end{align}
\end{subequations}
respectively be the real-time upward reserve deployed by generator $i$, the  real-time downward reserve deployed by generator $i$, the real-time VRES power spill by the VRES generator in bus $n$, and the real-time curtailment of load $j$. In equation~\eqref{eq:controls}, the variables $r_i^u, r_i^d, w_n^{\spill}, s_j$ are now respectively interpreted as the nominal (i.e., when there are no errors in the VRES power generation forecast) upward reserve deployed by generator $i$, downward reserve deployed by generator $i$, VRES power spill by the VRES generator in bus $n$, and curtailment of load $j$, for all $n \in \B$, $i \in \I$, and $j \in \J$. Furthermore, the variables $\alpha_i^u, \alpha_i^d, \beta_n, \gamma_j \ge 0$ define the affine controls being used in real-time for all  
$n \in \B$, $i \in \I_n$, $j \in \J_n$. Note that because for any $n\in \B$, the distribution of $\Delta  \W_{n}^f$ is symmetric and has zero mean, the choice of the $\Delta \W_n^f$ terms' signs in~\eqref{eq:controls} is done without loss of generality (see a more detailed discussion in Appendix~\ref{sec:assumptions}).

Note that when the uncertainty in the VRES power generation is considered, the nominal real-time balance constraint~\eqref{eq:deter_reblanace} for any bus 
$n \in \B$
is now given by
\begin{equation}
\label{eq:uncertainrebalance}
\sum_{i \in \I_n}(\r_i^u - \r_i^d) + \sum_{j \in \J_n} \s_j + \W_n- w_n^{\sched}  - \w_n^{\spill}+ \sum_{(n,l) \in \L} B_{nl} (\delta_{n}^0 - \delta_{n} - \delta_{l}^0 + \delta_{l}) = 0.
\end{equation}
After replacing~\eqref{eq:controls} into~\eqref{eq:uncertainrebalance}, one obtains that~\eqref{eq:uncertainrebalance} is equivalent to
\begin{equation}
\label{eq:uncertainrebalance2}
\begin{split}
\sum_{i \in \I_n}(r_i^u - r_i^d) + \sum_{j \in \J_n} s_j + W_n^f - w_n^{\sched} -w_n^{\spill} + \sum_{(n,l) \in \L} B_{nl} (\delta_{n}^0 -\delta_{n} - \delta_{l}^0 + \delta_{l}) + \\ \Delta \W_n^f \left(1 - \sum_{i \in \I_n}(\alpha_i^u + \alpha_i^d) - \sum_{j \in \J_n}\gamma_j  - \beta_n \right ) = 0,
\end{split}
\end{equation}
for all 
$n \in \B$.
Thus, to ensure that the network balance is maintained in real-time, when there is uncertainty in the VRES power generation is consided, the following constraint is imposed on the affine control variables in~\eqref{eq:controls}.
\begin{equation}
\label{eq:affine}
\sum_{i \in \I_n}(\alpha_i^u+ \alpha_i^d) + \sum_{j \in \J_n}\gamma_j + \beta_n= 1,
\end{equation}
for all 
$n \in \B$. 
That is, by combining equations~\eqref{eq:affine} and~\eqref{eq:uncertainrebalance2}, the real-time balance constraint is modeled by the market administrator as a ``hard'' constraint that cannot be violated. As discussed in~\citet[][Section 1.3]{bienstock2014chance}, both affine controls and the constraint on the affine controls reflect the way in which generator output is modulated, in real-time, in response to demand fluctuations.

Now, we are ready to present the CCO model of the electricity market taking into account the uncertainty in VRES power generation. Namely, 
	\begin{subequations}\label{eq:CCmodel}
		\begin{align}
		\min & ~\mathbb{E} \left (\sum_{i \in \I}(C_i p_i + C_i^u \r_i^u - C_i^d \r_i^d)+ \sum_{n \in \B} C_n^w (\W_n - \w_n^{\spill}) + \sum_{j \in \J} V_j\s_j \right ) \label{eq:uncer_obj} \\
		\st   & \sum_{i \in \I_n} p_i + w_n^{\sched} - \sum_{(n,l) \in \L} B_{nl} (\delta_{n}^0 -  \delta_{l}^0) =\sum_{j \in \J_n}L_j, \label{eq:uncer_balance} & \forall n\in \B\\
		& \sum_{i \in \I_n}(r_i^u - r_i^d) + \sum_{j \in \J_n} s_j + (W_n^f - w_n^{\sched}  - w_n^{\spill})+ \nonumber \\
		& \hspace{2in} \sum_{(n,l) \in \L} B_{nl} (\delta_{n}^0 - \delta_{n} - \delta_{l}^0 + \delta_{l}) = 0,   \label{eq:uncer_reblanace} & \forall n\in \B  \\
		& \sum_{i \in \I_n}(\alpha_i^u + \alpha_i^d) + \sum_{j \in \J_n}\gamma_j + \beta_n= 1,  \label{eq:controlconst} & \forall n \in \B \\
		& B_{kl}(\delta^0_{k} - \delta^0_{l}) \leq   \overline{C}_{kl}, \label{eq:uncer_line1} & \forall (k,l) \in \L \\ 
		& B_{kl}(\delta_{k}-\delta_{l}) \leq   \overline{C}_{kl}, \label{eq:uncer_line2} & \forall (k,l) \in \L   \\
		&  0 \le w_n^{\sched} \leq \overline{W}_n, \label{eq:uncer_wind_limit} &  \forall n \in \B     \\ 
		&  0 \le p_i  \leq \overline{P}_i,\label{eq:uncer_con_limit} & \forall i \in \I    \\
		& \mathbb{P}(\w_n^{\spill} \ge 0) \ge 1-\epsilon, ~\mathbb{P}(\w_n^{\spill} \le \W_n)\ge 1-\epsilon,\label{eq:uncer_spill_wind_limit} & \forall n \in \B    \\
		& \mathbb{P}(\r_i^u \ge 0) \ge 1-\epsilon, ~\mathbb{P}(\r_i^u \le \overline{R}_i^u)\ge 1-\epsilon,\label{eq:uncer_rampup_limit} & \forall i \in \I    \\
		& \mathbb{P}(\r_i^d \ge 0) \ge 1-\epsilon, ~\mathbb{P}(\r_i^d \le \overline{R}_i^d)\ge 1-\epsilon,\label{eq:uncer_rampdown_limit} & \forall i \in \I    \\
		& \mathbb{P}(p_i +  \r_i^u - \r_i^d \ge 0)\ge 1-\epsilon,  ~\mathbb{P}(p_i + \r_i^u - \r_i^d \le \overline{P}_i)\ge 1-\epsilon,  \label{eq:uncer_con2_limit} & \forall i \in \I     \\
		& \mathbb{P}(\s_j \ge 0) \ge 1-\epsilon, ~ \mathbb{P}(\s_j \le L_j) \ge 1-\epsilon, \label{eq:uncer_demand_limit} & \forall j \in \J    \\
		& \alpha_i^u, \alpha_i^d, p_i,  r_i^u, r_i^d \ge 0, & \forall i \in \I,\\
				&w_n^{\sched},w_n^{\spill}, \beta_n, \gamma_j, s_j \geq 0, & \forall n \in \B, j \in \J
		\end{align}
	\end{subequations}
where $\mathbb{P}(\cdot)$ indicates the probability of an event, and $0 <\epsilon < 0.5$ is the tolerance for constraint violations in the electricity market. Note that similar to the real-time balance constraint, we consider ``hard'' line capacity constraints (i.e.,~\eqref{eq:uncer_line1},~\eqref{eq:uncer_line2}); that is, no violations to the line capacity constraints in real-time are allowed. Throughout the article, we use the following blanket assumptions on problem~\eqref{eq:CCmodel}.

\begin{assumption}
\label{as:blanket}
In what follows, it is always assumed that:
\begin{enumerate}[label=(\roman*)]
\item The CCO market model~\eqref{eq:CCmodel} is feasible.
\label{as:blanket1}
\item $\sum_{j \in \J}(L_j - s^*_j) > 0$, where $s_j^* = \argmin_{s_j}\{\eqref{eq:CCmodel}\}$, for all $j \in \J$.
\label{as:blanket2}
\end{enumerate}
\end{assumption}

Note that Assumption~\ref{as:blanket}\ref{as:blanket2} is a mild assumption; namely, Assumption~\ref{as:blanket}\ref{as:blanket2} is equivalent to assuming that the naive solution of not dispatching any power is not optimal for the CCO market-clearing model~\eqref{eq:CCmodel}. For example, this assumption is satisfied simply if the cost associated with load curtailment (i.e., the value of lost load~\cite{woo1991, lawton2003}) $V_j$ is large enough for one $j \in \J$. To obtain a {\em deterministic equivalent} of the CCO market-clearing model~\eqref{eq:CCmodel}, we use the fact that~\citep[see, e.g.,][]{nemirovski2006convex} the deterministic equivalent of the constraint $\mathbb{P}(a{\mathbf{x}}  \le b) \ge 1- \epsilon$ (resp., $\mathbb{P}(a{\mathbf{x}}  \ge b) \ge 1- \epsilon$) for a random variable ${\mathbf{x}} \sim \mathcal{N}(0, \sigma)$, $a \ge 0$, $b \in \R$  is given by the constraint $b - a\Phi^{-1}_{1-\epsilon}\sigma \ge 0$ (resp., $b + a\Phi^{-1}_{1-\epsilon}\sigma \le 0$), where for any $0 < \delta <1$, $\Phi^{-1}_{\delta}$ is the $\delta$-quantile of the standard normal distribution. Note these deterministic equivalents hold when $\sigma = 0$, as in that case, $\mathbf{x}=0$, since it is assumed that~$\mathbb{E}(\mathbf{x}) = 0$. Using this result, after replacing~\eqref{eq:controls} in~\eqref{eq:CCmodel}, and noticing that the fact that $\mathbb{E}(\Delta \W_n^f) = 0$ for all $n \in \B$ implies that the expectation in~\eqref{eq:uncer_obj} is equivalent to 
$\sum_{i \in \I}(C_i p_i + C_i^u r_i^u - C_i^d r_i^d)+ \sum_{n \in \B} C_n^w (W_n^f - w_n^{\spill}) + \sum_{j \in \J} V_js_j$, 
it follows that the deterministic equivalent of the CCO market model~\eqref{eq:CCmodel} is given by: 
	\begin{subequations}\label{eq:detCCmodel}
		\begin{align}
		\min & \sum_{i \in \I}(C_i p_i + C_i^u r_i^u - C_i^d r_i^d)+ \sum_{n \in \B} C_n^w (W_n^f - w_n^{\spill})+ \sum_{j \in \J} V_js_j \label{eq:detuncer_obj} \\
		\st   & \sum_{i \in \I_n} p_i + w_n^{\sched} - \sum_{(n,l) \in \L} B_{nl} (\delta_{n}^0 -  \delta_{l}^0) =\sum_{j \in \J_n}L_j , \label{eq:detuncer_balance}  & \hspace{-2em}(\lambda_n)&& \forall n\in \B\\
		& \sum_{i \in \I_n}(r_i^u - r_i^d) + \sum_{j \in \J_n} s_j + (W_n^f - w_n^{\sched}  - w_n^{\spill}) + \nonumber \\
		& \hspace{2in}  \sum_{(n,l) \in \L} B_{nl} (\delta_{n}^0 - \delta_{n} - \delta_{l}^0 + \delta_{l}) = 0,   \label{eq:detuncer_reblanace} &\hspace{-2em}(\nu_n)&& \forall n\in \B  \\
		& \sum_{i \in \I_n}(\alpha_i^u + \alpha_i^d) + \sum_{j \in \J_n}\gamma_j + \beta_n= 1,  \label{eq:detcontrolconst} &\hspace{-2em}(\kappa_n)&& \forall n \in \B \\
		& B_{kl}(\delta^0_{k} - \delta^0_{l}) \leq   \overline{C}_{kl}, \label{eq:detuncer_line1} &\hspace{-2em}&& \forall (k,l) \in \L \\ 
		&  B_{kl}(\delta_{k}-\delta_{l}) \leq   \overline{C}_{kl}, \label{eq:detuncer_line2} &\hspace{-2em}&& \forall (k,l) \in \L   \\
		&  - w_n^{\sched}\ge  -\overline{W}_n, \label{eq:detuncer_wind_limit} &\hspace{-2em}(\mu_n)&&  \forall n \in \B     \\ 
		&   -p_i \ge -\overline{P}_i,\label{eq:detuncer_con_limit} &\hspace{-2em}(\rho_i)&& \forall i \in \I    \\
		& w_n^{\spill} - \beta_n \Phi^{-1}_{1-\epsilon} \sigma_{n} \ge 0, 
		~-w_n^{\spill} - (1-\beta_n)  \Phi^{-1}_{1-\epsilon} \sigma_{n}\ge -W_n^f ,\label{eq:detuncer_spill_limit} &\hspace{-2em}(y_n^{\spill}, x_n^{\spill})&& \forall n \in \B    \\
		& r_i^u - \alpha_i^u \Phi^{-1}_{1-\epsilon} \sigma_{b(i)} \ge 0, 
		~-r_i^u - \alpha_i^u  \Phi^{-1}_{1-\epsilon} \sigma_{b(i)}\ge -\overline{R}_i^u,\label{eq:detuncer_rampup_limit} &\hspace{-2em}(y_i^u, x_i^u)&& \forall i \in \I    \\
		& r_i^d - \alpha_i^d \Phi^{-1}_{1-\epsilon} \sigma_{b(i)} \ge 0, 
		~-r_i^d - \alpha_i^d  \Phi^{-1}_{1-\epsilon} \sigma_{b(i)}\ge -\overline{R}_i^d,\label{eq:detuncer_rampdown_limit} &\hspace{-2em}(y_i^d, x_i^d)&&\forall i \in \I    \\
		& p_i +  r_i^u - r_i^d -(\alpha_i^u+\alpha_i^d)\Phi^{-1}_{1-\epsilon} \sigma_{b(i)} \ge 0,  \label{eq:detuncer_con1_limit} &\hspace{-2em}(y_i)&& \forall i \in \I     \\
				& -p_i - r_i^u+ r_i^d - (\alpha_i^u + \alpha_i^d)\Phi^{-1}_{1-\epsilon} \sigma_{b(i)} \ge -\overline{P}_i,  \label{eq:detuncer_con2_limit} &\hspace{-2em}(x_i)&& \forall i \in \I     \\
		& s_j - \gamma_j \Phi^{-1}_{1-\epsilon} \sigma_{b(j)} \ge 0, 
		~-s_j - \gamma_j  \Phi^{-1}_{1-\epsilon} \sigma_{b(j)}\ge -L_j, \label{eq:detuncer_demand_limit} &\hspace{-2em}(y_i^s, x_i^s)&& \forall j \in \J    \\
		&\alpha_i^u, \alpha_i^d, p_i,  r_i^u, r_i^d \ge 0, \label{eq:nonneg1}&&& \forall i \in \I\\
		&w_n^{\sched}, w_n^{\spill}, \beta_n, \gamma_j, s_j \geq 0. \label{eq:nonneg2}&&& \forall n \in \B, j \in \J		\end{align}
	\end{subequations}

The deterministic CCO (DCCO) market model~\eqref{eq:detCCmodel} is a linear program (LP). In problem~\eqref{eq:detCCmodel}, except for the line capacity constraints~\eqref{eq:detuncer_line1},~\eqref{eq:detuncer_line2}, and the non-negative constraints~\eqref{eq:nonneg1},~\eqref{eq:nonneg2}, the dual variables associated with each of the constraints are labeled in parenthesis at the right of the constraint (e.g., the dual variable associated with the real-time balance constraint~\eqref{eq:detuncer_reblanace} at bus $n \in \B$ is $\nu_n$). Note that all the inequalities, except for the lower bounds in constraints~\eqref{eq:detuncer_line1},~\eqref{eq:detuncer_line2} are expressed as greater than or equal constraints. This is done for simplicity, making the dual variables~$\kappa_n, \mu_n, \rho_i, y_n^{\spill}, x_n^{\spill}, y_i^u, x_i^u, y_i^d, x_i^d, y_i, x_i, y_i^s, x_i^s$ non-negative for all $n \in \B, i \in \I, j \in \J$~\citep[see, e.g.,][]{chvatal1983linear}. Similar to~\citet{morales2012pricing} and related articles~\citep[see, e.g.,][among others]{ref.oneill2005, ref.ruiz2012pricing, bjorndal2008equilibrium, kuang2019pricing}, the optimal values of these dual variables are used to define the proposed revenue adequate pricing scheme in what follows.

\section{Revenue Adequate Pricing Scheme}
\label{sec:CCPricing}

We now present the main result of the article; that is, a pricing scheme for the CCO electricity market-clearing model~\eqref{eq:CCmodel} that ensures that the power generators in the market and the market administrator attain revenue adequacy (or cost recovery) in expectation.

In what follows, for any primal or dual variable $(\cdot)$ of the CCO market model~\eqref{eq:CCmodel} or DCCO market model~\eqref{eq:detCCmodel}, $(\cdot)^*$ indicates the optimal value of the decision variable (e.g., $p_i^*$, $\lambda_n^*$ are respectively the optimal value of $p_i$, the scheduled power generation of generator $i \in \I$, and the optimal value of $\lambda_n$, the dual variable associated with the power balance constraint at bus $n \in \B$ in the scheduling stage). Now, for any $i \in \I_n$, $n\in \B$, let
\begin{equation}
\label{eq:tau}
\tau_i^u = \left \{\begin{array}{ll}
			 	\dfrac{\kappa_{n}^*}{{\sigma_n} \Phi_{1 - \epsilon}^{-1} } & \text{ if $\kappa_{n}^* - {\sigma_n} \Phi_{1 - \epsilon}^{-1} {y^u_i}^* \ge 0$}\\
				{y^u_i}^* & \text{ if $\kappa_{n}^* - {\sigma_n} \Phi_{1 - \epsilon}^{-1} {y^u_i}^* \le 0$} \\
				\end{array} 
				\right., \qquad
				\tau_i^d = \left \{ \begin{array}{ll}
			 	\dfrac{\kappa_{n}^*}{{\sigma_n} \Phi_{1 - \epsilon}^{-1} } & \text{ if $\kappa_{n}^* - {\sigma_n} \Phi_{1 - \epsilon}^{-1} {y^d_i}^* \ge 0$}\\
				{y^d_i}^* & \text{ if $\kappa_{n}^* - {\sigma_n} \Phi_{1 - \epsilon}^{-1} {y^d_i}^* \le 0$}\\
				\end{array} 
				\right.,
\end{equation}
and 
\begin{equation}
 \label{def:kappa}
\ccozeta = \frac{\dsum_{n \in \B}  \left (\dsum_{i \in \I_n }(\tau_i^u{r_i^u}^* +\tau_i^d{r_i^d}^* ) -({y_n^{\spill}}^*- {x_n^{\spill}}^*)(W_n^f - {w_n^{\spill}}^*)\right)}{\dsum_{n \in \B} \sum_{j \in \J_n} (L_j -  {s_j^*})}.
\end{equation}
Note that the assumptions $\sigma_n > 0$, for all $n \in \B$; $0 < \epsilon < 0.5$, implying $\Phi^{-1}_{1-\epsilon} > 0$; and Assumption~\ref{as:blanket}\ref{as:blanket2}, ensure that $\tau_i^u, \tau_i^d$, for all $i \in \I_n$, $n \in \B$, and $\zeta$ are well defined.

Below, we introduce the pricing scheme for the CCO market-clearing model~\eqref{eq:CCmodel}. 

\begin{definition}[Chance-Constrained Optimization Pricing Scheme] 
\label{def:CCpricing} Energy transactions settled via the CCO market-clearing model~\eqref{eq:CCmodel}
are priced as follows:
\begin{enumerate}[label =(\roman*)]
	\item Each conventional generator $i \in \I_n, n\in \B$, is compensated at price
	$\lambda_{n}^*$ 
	for every unit of scheduled power ${p_i}^*$. VRES generators at bus $n \in \B$ are compensated at price $\lambda_n^*-{y_n^{\spill}}^* + {x_n^{\spill}}^*$ for every unit of scheduled power~${w_n^{\sched}}^*$.
	\label{it:CCpricegens}
	\item Each load $j \in \J_n, n \in \B$ is charged at price $\lambda_n^* + \ccozeta$ for every unit of its scheduled consumption~${L_j}$.
	\item Each conventional generator $i \in \I_n, n\in \B$ is compensated 
	at price $\nu_n^*+\tau^u_i$ (resp., charged at price $\nu_n^*-\tau^d_i$) 
	for every unit of real-time upward reserve  deployed  ${\r^u_i}^*$ (resp., downward reserve deployed ${\r^d_i}^*$).
	\label{it:CCpriceloads}
	\item VRES generators at bus $n\in \B$ are compensated (resp., charged) at price $\nu_n^* -{y_n^{\spill}}^* + {x_n^{\spill}}^*$ for every unit of  surplus (resp., shortage) of real-time net VRES power generated over (resp., under) the VRES power generation scheduled $\max\{0,\W_n - {w_n^{\sched}}^* - {\w_n^{\spill}}^*\}$ (resp., $\max\{0,-\W_n + {w_n^{\sched}}^*+ {\w_n^{\spill}}^*\}$). \label{it:CCpricewinds} \looseness -1
	\item Each load $j\in\J_n, n \in \B$ is compensated at price $\nu_n^*+ \ccozeta$ for every unit of real-time involuntary load curtailment $\s_j^*$.
	\label{it:CCpricesheds}
\end{enumerate}
\end{definition}

\looseness=-1
As formally stated below, the CCO pricing scheme in Definition~\ref{def:CCpricing} guarantees that revenue adequacy (or cost recovery) in expectation is satisfied for the power generators and for the market administrator in the CCO market-clearing model~\eqref{eq:CCmodel}. The proof that these properties are satisfied is presented in Sections~\ref{sec:ISOrev},~\ref{sec:Gencost}, and~\ref{sec:Windcost}, using a combination of LP and Lagrangian duality techniques similar to the ones used in~\citet{ref.oneill2005, morales2012pricing, kuang2018pricing}.

\begin{theorem}
\label{thm:CCpricing}
The CCO pricing scheme introduced in Definition~\ref{def:CCpricing} guarantees that in the CCO market-clearing model~\eqref{eq:CCmodel}, the market administrator is ensured revenue adequacy in expectation (i.e., has non-negative profit in expectation), and the generating units in the market, including the VRES generators, are ensured cost recovery in expectation (i.e., their revenue is greater than or equal than their operating costs in expectation).
\end{theorem}

\proof{Proof.} The statement follows from the discussion in Section~\ref{sec:ISOrev} (revenue adequacy of the market administrator), Section~\ref{sec:Windcost} (cost recovery of VRES generators), and
Section~\ref{sec:Gencost} (cost recovery of conventional generators). \hfill \qed
\endproof

\subsection{Revenue Adequacy of Market Administrator}
\label{sec:ISOrev}
From problem~\eqref{eq:CCmodel} and Definition~\ref{def:CCpricing}, it follows that the expected profit for the market administrator in the CCO market-clearing model~\eqref{eq:CCmodel} is equal to:
\begin{equation}
\label{eq:ISOexprevenue}
\begin{array}{lcll}
\Gamma^o &= & \left .\displaystyle \dsum_{n \in  \B} \left (- \dsum_{i \in \I_n}\lambda_n^*p_i^*-  (\lambda_n^*-{y_n^{\spill}}^* + {x_n^{\spill}}^*) {w_n^{\sched}}^* + \dsum_{j \in \J_n}(\lambda_n^* + \ccozeta)L_j \right) 
 \right \} \Gamma^o_{\rm scheduled}\\ 
		&& \left . \begin{array}{l} \displaystyle -  \dsum_{n \in \B}  \Bigg (\dsum_{i \in \I_n} \Big((\nu_n^* + \tau^u_i)\mathbb{E}({\r_i^u}^*) - (\nu_n^* - \tau^d_i)\mathbb{E}({\r_i^d}^*) \Big) + \\
		(\nu_n^* -{y_n^{\spill}}^* + {x_n^{\spill}}^*) \mathbb{E}( {\W}_n - {w_n^{\sched}}^*- {\w_n^{\spill}}^*) +\dsum_{j \in \J_n}(\nu_n^*  + \ccozeta)\mathbb{E}(\s_j^*)  \Bigg ),
		\end{array} \right \} \Gamma^o_{\rm real-time}
\end{array}
\end{equation}
where we have used the linearity of the expectation, and the fact that for any $a \in \R$, $\max\{0,a\}-\max\{0,-a\} = a$ to rewrite the VRES generator's real-time compensation.

 In~\eqref{eq:ISOexprevenue}, $\Gamma_{\rm schedule}^o$ is the scheduled profit; that is, 
$\sum_{n \in  \B} (  \sum_{i \in \I_n}\lambda_n^*p_i^* + (\lambda_n^*-{y_n^{\spill}}^* + {x_n^{\spill}}^*){w_n^{\sched}}^*)$ 
is the cost that the market administrator needs to pay for the scheduled generation of both conventional and VRES generators, and  $\sum_{n \in  \B} \sum_{j \in \J_n}(\lambda_n^* + \ccozeta)L_j$ is the revenue that the market administrator obtains from the scheduled loads. Furthermore, $\Gamma_{\rm real-time}^o$ is the real-time expected profit; that is,
		$\sum_{n \in \B}\sum_{i \in \I_n}(\nu_n^* + \tau^u_i)\mathbb{E}({\r_i^u}^*)$ (resp., $\sum_{n \in \B}\sum_{i \in \I_n}(\nu_n^* - \tau^d_i)\mathbb{E}({\r_i^d}^*$))
		is the expected cost (resp., revenue) that the market administrator pays (resp., receives) for upward reserve deployments (resp., downward reserve deployments) in conventional generation in the real-time stage; $\sum_{n \in \B} (\nu_n^* -{y_n^{\spill}}^* + {x_n^{\spill}}^*)  \mathbb{E}({\W}_n-  {w_n^{\sched}}^* - {\w_n^{\spill}}^*)$ 
		 is the expected cost (resp., revenue) that the market administrator pays (resp., receives) for the surpluses (resp., shortages) of real-time net VRES power generated over (resp., under) the VRES power generation scheduled $\max\{0, {\W}_n-  {w_n^{\sched}}^*- {\w_n^{\spill}}^*\}$ (resp., $\max\{0, -{\W}_n+  {w_n^{\sched}}^*+ {\w_n^{\spill}}^*\}$) for all $n \in \B$; and $\sum_{n \in  \B} \sum_{j \in \J_n} (\nu_n^*  + \ccozeta)\mathbb{E}(s_j^*)$ is the  expected cost that the market administrator pays for the real-time involuntary load curtailments.
		 
After using~\eqref{eq:winddef} and~\eqref{eq:controls}, to compute the expectations in~\eqref{eq:ISOexprevenue}, one gets, after rearranging terms, that:
\begin{equation}
\label{eq:detISOexprevenue}
\begin{array}{lcl}
\Gamma^o &= &\displaystyle \underbrace{
\dsum_{n \in  \B} \lambda_n^* \left (\dsum_{j \in \J_n} L_j - \left(\dsum_{i \in \I_n} p_i^* + {w_n^{\sched}}^* \right)\right )
-\dsum_{n \in  \B} \nu_n^*\left (\dsum_{i \in \I_n} ({r_i^u}^* - {r_i^d}^*) +\dsum_{j \in \J_n} s_j^*+ (W_n^f - {w_n^{\sched}}^* - {w_n^{\spill}}^*)  \right )
}_{\Gamma^o_1} \\ 
&  & \displaystyle \underbrace{
-\dsum_{n \in  \B} \left (\dsum_{i \in \I_n} (  \tau^u_i{r_i^u}^* + \tau^d_i{r_i^d}^*) -({y_n^{\spill}}^*- {x_n^{\spill}}^*)(W_n^f - {w_n^{\spill}}^*) \right ) + \ccozeta \dsum_{n \in  \B}\dsum_{j \in \J_n}(L_j - s_j^*).
}_{\Gamma^o_2} \\ 
\end{array}
\end{equation}
Note that from the definition of $\ccozeta$ in~\eqref{def:kappa}, it follows that $\Gamma_2^o = 0$ in~\eqref{eq:detISOexprevenue}. Thus, to show that the market administrator's expected profit in the CCO market-clearing model~\eqref{eq:CCmodel} is non-negative, we next show that $\Gamma^o_1 \ge 0$.

From~\eqref{eq:detuncer_balance}, and~\eqref{eq:detuncer_reblanace}, it follows that:
\begin{subequations}
\begin{align}
&\lambda_n^* \left (\sum_{j \in \J_n}L_j - \left (\sum_{i \in \I_n} p_i^* + {w_n^{\sched}}^* - \sum_{(n,l) \in \L} B_{nl} ({\delta_{n}^0}^* -  {\delta_{l}^0}^*) \right) \right)= 0 & \forall n \in \B, \label{eq:comp1}\\
&\nu_n^*\Bigg (- \Bigg (\sum_{i \in \I_n}({r_i^u}^* + {r_i^d}^*) + \sum_{j \in \J_n} s_j^* + (W_n^f - {w_n^{\sched}}^*  - {w_n^{\spill}}^*) + \nonumber\\ 
& \hspace{2in} \sum_{(n,l) \in \L} B_{nl} ({\delta_{n}^0}^* - {\delta_{n}}^* - {\delta_{l}^0}^* + {\delta_{l}}^*) \Bigg) \Bigg)= 0 & \forall n \in \B. \label{eq:comp2}
\end{align}
\end{subequations}

After adding~\eqref{eq:comp1} and~\eqref{eq:comp2}, and then taking the sum over all $n \in \B$, it follows that:
\begin{equation}
\label{eq:newgammao1}
\Gamma^o_1 = \dsum_{n \in \B}\left ( -\lambda_n^* \sum_{(n,l) \in \L} B_{nl} ({\delta_{n}^0}^* -  {\delta_{l}^0}^*) +\nu_n^*\sum_{(n,l) \in \L} B_{nl} ({\delta_{n}^0}^* - \delta_{n}^* - {\delta_{l}^0}^* + \delta_{l}^*) \right).
\end{equation}
Now consider the following LP:
\begin{equation}
\label{eq:deltasub}
\begin{array}{lllll}
z^o = &\min & \dsum_{n \in \B}\left ( \lambda_n^* \sum_{(n,l) \in \L} B_{nl} ({\delta_{n}^0} -  {\delta_{l}^0}) -\nu_n^*\sum_{(n,l) \in \L} B_{nl} ({\delta_{n}^0} - \delta_{n} - {\delta_{l}^0} + \delta_{l}) \right)\\
		& \st &  B_{kl}(\delta^0_{k} - \delta^0_{l}) \leq   \overline{C}_{kl}, & \forall (k,l) \in \L \\ 
		& &  B_{kl}(\delta_{k}-\delta_{l}) \leq   \overline{C}_{kl}. & \forall (k,l) \in \L   \\		
		\end{array}
		\end{equation}
Note that 	$\delta_{n}^0 = \delta_{n} = 0$, for all $n \in \B$ is a feasible solution of~\eqref{eq:deltasub}. Thus, $z^o \le 0$. Furthermore, we claim that $z^o = -\Gamma^o_1$. Thus, $\Gamma^o_1 \ge 0$. After replacing $\Gamma^o_1 \ge 0$, $\Gamma^o_2 = 0$ in~\eqref{eq:detISOexprevenue}, it follows that the pricing scheme introduced in Definition~\ref{def:CCpricing} guarantees that in the CCO market-clearing model~\eqref{eq:CCmodel}, the market administrator has non-negative profit in expectation.

The claim that $z^o = -\Gamma^o_1$ follows by noticing that the LP~\eqref{eq:deltasub} is the subproblem on the decision variables $\delta_{n}^0, \delta_{n}$, for all $n \in \B$, in which the Lagrangean relaxation of~\eqref{eq:detCCmodel} with respect to constraints~\eqref{eq:detuncer_balance}, and~\eqref{eq:detuncer_reblanace} decomposes after replacing $\lambda_n \to \lambda_n^*, \nu_n \to \nu_n^*$, for all $n \in \B$ in the Lagrangean relaxation. Then, from LP and Lagrangean duality (in particular, from {\em saddle point} results~\citep[see, e.g.,][Sec.~5.4.2]{boyd2004convex}), it follows that the optimal value of~\eqref{eq:deltasub} is obtained after replacing $\delta_{n}^0 \to {\delta_n^0}^*, \delta_{n} \to \delta_n^*$, for all $n \in \B$ in the objective function of~\eqref{eq:deltasub}.

For the purpose of numerical analysis of the CCO pricing scheme introduced in Definition~\ref{def:CCpricing}, it is useful to present the standard deviation of the market administrator's profit $\sigma^o$; namely, from~\eqref{eq:winddef}, \eqref{eq:controls}, and Definition~\ref{def:CCpricing}, it follows, under the additional assumption of $\Delta \W_n^f$ being independently distributed for all $n \in \B$, that
\begin{equation}
\label{eq:operstd}
\sigma^o = \sqrt{\dsum_{n \in \B} \sigma_n^2 \left (\sum_{i \in \I_n} \left ({\alpha_i^u}^*(\nu_n^* + \tau_i^u) + 
{\alpha_i^d}^*(\nu_n^* - \tau_i^d) \right) - (\nu_n^* - {y_n^{\spill}}^* + {x_n^{\spill}}^*)(1-\beta_n^*) + 
\sum_{j \in \J_n} \gamma_j^*(\nu_n^* +\ccozeta)  \right )^2}.
\end{equation}

\subsection{Cost Recovery of VRES Generators}
\label{sec:Windcost}		

From problem~\eqref{eq:CCmodel} and Definition~\ref{def:CCpricing}, it follows that the VRES generators' expected profit at bus~$n \in \B$ in  the CCO market-clearing model~\eqref{eq:CCmodel} is equal to:
\begin{equation}
\label{eq:WINDexprevenue}
\begin{split}
\Gamma^{w_n} = (\lambda_n^*-{y_n^{\spill}}^* + {x_n^{\spill}}^*){w_n^{\sched}}^* + (\nu_n^* -{y_n^{\spill}}^* + {x_n^{\spill}}^*)\mathbb{E}(\W_n - {w_n^{\sched}}^* - {\w_n^{\spill}}^*) - C_n^w\mathbb{E}(\W_n-{\w_n^{\spill}}^*),
\end{split}
\end{equation}
where we have used the linearity of the expectation, and the fact that for any $a \in \R$, $\max\{0,a\}-\max\{0,-a\} = a$ to rewrite the VRES generator's real-time compensation.

In~\eqref{eq:WINDexprevenue}, $(\lambda_n^*-{y_n^{\spill}}^* + {x_n^{\spill}}^*) {w_n^{\sched}}^*$ is the VRES generators' revenue for scheduled VRES power; $(\nu_n^* -{y_n^{\spill}}^* + {x_n^{\spill}}^*)  \mathbb{E}({\W}_n-  {w_n^{\sched}}^* - {\w_n^{\spill}}^*)$ 
		 is the expected cost (resp., revenue) that the market administrator pays (resp., receives) for the surplus (resp., shortage) of real-time net VRES power generated over (resp., under) the VRES power generation scheduled $\max\{0, {\W}_n-  {w_n^{\sched}}^*- {\w_n^{\spill}}^*\}$ (resp., $\max\{0, -{\W}_n+  {w_n^{\sched}}^*+ {\w_n^{\spill}}^*\}$); and $C_n^w\mathbb{E}(\W_n-{\w_n^{\spill}}^*)$ is the expected VRES generators' cost for the net real-time VRES power generation.

After using~\eqref{eq:winddef} and~\eqref{eq:controlssp}, to compute the expectations in~\eqref{eq:WINDexprevenue}, and simplifying, one gets that:
\begin{equation}
\label{eq:detWINDexprevenue}
\Gamma^{w_n} = (\lambda_n^* - \nu_n^*) {w_n^{\sched}}^*+(C_n^w - \nu_n^* +{y_n^{\spill}}^* - {x_n^{\spill}}^*)  {w_n^{\spill}}^*+(\nu_n^* -{y_n^{\spill}}^* + {x_n^{\spill}}^*-C_n^w)W_n^f.
\end{equation}
To continue, we need to introduce the dual complementary conditions associated with the variables $w_n^{\sched}, w_n^{\spill}$, for all $n \in \B$ of the LP~\eqref{eq:detCCmodel},
\begin{equation}
\label{eq:windcomplementary}
\begin{array}{ll}
{w_n^{\sched}}^*(-\lambda_n^* + \nu_n^* + \mu_n^*) = 0, & \forall n \in \B,\\
{w_n^{\spill}}^*(-C_n^w + \nu_n^* -{y_n^{\spill}}^*+{x_n^{\spill}}^*) = 0, & \forall n \in \B,\\
\end{array}
\end{equation}
as well as the dual constraint associated with the variable $w_n^{\sched}$, for all $n \in \B$ of the LP~\eqref{eq:detCCmodel},
\begin{equation}
\label{eq:windconstraint}
-C_n^w + \nu_n^* -{y_n^{\spill}}^*+{x_n^{\spill}}^* \ge 0, \forall n \in \B,\\
\end{equation}
Then it follows that for any $n \in \B$, \eqref{eq:detWINDexprevenue} is equivalent to:
\begin{equation}
\label{eq:detWINDexprevenue2}
\Gamma^{w_n} = \mu_n^* {w_n^{\sched}}^*+(-C_n^w+\nu_n^* -{y_n^{\spill}}^* + {x_n^{\spill}}^*)W_n^f \ge 0,
\end{equation} 
where the equality follows from using~\eqref{eq:windcomplementary}, and the inequality follows from~\eqref{eq:windconstraint}, and the fact that $\mu_n^*$, $w_n^{\sched}$, $W_n^f \ge 0$, for all $n \in \B$. 

Thus, $\Gamma^{w_n} \ge 0$; that is, the pricing scheme introduced in Definition~\ref{def:CCpricing} guarantees that in the CCO market-clearing model~\eqref{eq:CCmodel}, the expected revenue of the VRES generators in the market at bus $n$,  is greater than or equal than their expected operating costs (i.e., cost recovery is  guaranteed in expectation), for all $n \in \B$.

For the purpose of numerical analysis of the CCO pricing scheme introduced in Definition~\ref{def:CCpricing}, it is useful to present the standard deviation of the profit for VRES generators $\sigma^{w_n}$; namely, from~\eqref{eq:winddef}, \eqref{eq:controlssp}, and Definition~\ref{def:CCpricing}, it follows that for all $n \in \B$,
\begin{equation}
\label{eq:windstd}
\sigma^{w_n} = (-C_n^w+ \nu_n^* -{y_n^{\spill}}^* + {x_n^{\spill}}^*)(1-\beta_n^*)\sigma_n.
\end{equation}

\subsection{Cost Recovery of Conventional Generators}
\label{sec:Gencost}
From problem~\eqref{eq:CCmodel} and Definition~\ref{def:CCpricing}, it follows that the expected profit of conventional generator~$i \in \I_n$, $n \in \B$ in  the CCO market-clearing model~\eqref{eq:CCmodel} is equal to:
\begin{equation}
\label{eq:GENexprevenue}
\Gamma^{g_i} = \lambda_n^*p_i^* + (\nu_n^* +\tau^u_i)\mathbb{E}({\r_i^u}^*)+ C_i^d\mathbb{E}({\r_i^d}^*) - (C_i p_i^* + C_i^u \mathbb{E}({\r_i^u}^*) + (\nu_n^* -\tau^d_i)\mathbb{E}({\r_i^d}^*)),  
\end{equation}
where  
$\lambda_n^*p_i^*$ (resp., $C_i {p_i}^*$) 
is the conventional generator's revenue (resp., cost) for scheduled power, 
$(\nu_n^* +\tau^u_i)\mathbb{E}({\r_i^u}^*)$ (resp., $C_i^u \mathbb{E}({\r_i^u}^*)$) 
is the  conventional generator's expected revenue (resp., cost) for real-time upward reserve deployed, and 
$C_i^d\mathbb{E}({\r_i^d}^*)$ (resp., $(\nu_n^* -\tau^d_i)\mathbb{E}({\r_i^d}^*)$)
is the conventional generator's expected cost saving (resp., cost) for real-time downward reserve deployed.

After using~\eqref{eq:winddef}, \eqref{eq:controlsrup}, and~\eqref{eq:controlsrdo}, to compute the expectations in~\eqref{eq:WINDexprevenue}, and rearranging terms, one gets that:
\begin{equation}
\label{eq:detGENexprevenue}
\Gamma^{g_i} = \underbrace{- C_i p_i^* - C_i^u {r_i^u}^*+ C_i^d{r_i^d}^*+ \lambda_n^*p_i^* + \nu_n^*({r_i^u}^*-{r_i^d}^*)}_{\Gamma^{g_i}_1} + \underbrace{\tau^u_i{r_i^u}^* +\tau^d_i{r_i^d}^*}_{\Gamma^{g_i}_2}.
\end{equation}
For ease of exposition, in what follows, let 
\[
\sigma'_i := \Phi^{-1}_{1-\epsilon} \sigma_{b(i)} > 0
\]
(cf., discussion after~\eqref{def:kappa}),  for any $i \in \I$.
To continue, we need to introduce the primal complementary conditions associated with constraints: 
\eqref{eq:detuncer_con_limit},
the upper bounds on $r^u_i$ in~\eqref{eq:detuncer_rampup_limit},
the upper bounds on $r^d_i$ in~\eqref{eq:detuncer_rampdown_limit},
\eqref{eq:detuncer_con1_limit}, and
\eqref{eq:detuncer_con2_limit}, of the LP~\eqref{eq:detCCmodel},
\begin{equation}
\label{eq:primalcomp}
\begin{array}{ll}
(-\overline{P}_i +p_i^*) \rho_i^* = 0 & \forall i \in \I    \\
(-\overline{R}_i^u+ {r_i^u}^* + {\alpha_i^u}^* \sigma'_i){x_i^u}^*= 0,& \forall i \in \I    \\
(-\overline{R}_i^d+{r_i^d}^* + {\alpha_i^d}^* \sigma'_i){x_i^d}^*=0,&\forall i \in \I    \\
(-p_i^* -  {r_i^u}^* + {r_i^d}^* +({\alpha_i^u}^*+{\alpha_i^d}^*)\sigma'_i)y_i^* = 0, & \forall i \in \I     \\
(-\overline{P}_i+p_i^* + {r_i^u}^*- {r_i^d}^* + ({\alpha_i^u}^* + {\alpha_i^d}^*)\sigma'_i)x_i^* = 0,& \forall i \in \I     \\
 \end{array}
 \end{equation}
as well the dual complementary conditions associated with the variables $p_i, r_i^u, r_i^d, \alpha_i^u, \alpha_i^d$ for all $i \in \I$ of the LP~\eqref{eq:detCCmodel},
\begin{equation}
\label{eq:dualcomp}
\begin{array}{ll}
p_i^*(C_i - \lambda_{b(i)}^* + \rho_i^* -  y_i^*+x_i^*) = 0, & \forall i \in \I\\
{r_i^u}^* (C_i^u - \nu_{b(i)}^*  - {y_i^u}^*+ {x_i^u}^* -y_i^*  + x_i^*  ) = 0,& \forall i \in \I\\
{r_i^d}^* (-C_i^d + \nu_{b(i)}^* - {y_i^d}^* + {x_i^d}^* + y_i^* - x_i^* ) = 0, & \forall i \in \I\\
{\alpha_i^u}^*(-  \kappa_{b(i)}^* + \sigma'_i ({{{y _i^u}^*} +x_i^u}^* +y_i^*+  x_i^* ) ) = 0, & \forall i \in \I\\
{\alpha_i^d}^*( - \kappa_{b(i)}^* + \sigma'_i ({y_i^d}^*   +{x_i^d}^* +y_i^* +  x_i^*  )) = 0. & \forall i \in \I\\
\end{array}
\end{equation}
Now, consider subtracting 
from $\Gamma^{g_i}_1$ (defined in~\eqref{eq:detGENexprevenue}) 
the left hand sides of all the primal complementary constraints associated with generator $i$ in~\eqref{eq:primalcomp}. After doing this, and rearranging terms, one gets that:
\begin{equation}
\label{eq:genpart1}
\begin{array}{lcl}
\Gamma_1^{g_i} & = & p_i^*(- C_i + \lambda_n^* - \rho_i^*+y_i^*-x_i^*) + \\
                             &      & {r_i^u}^*(- C_i^u+ \nu_n^* - {x_i^u}^* + y_i^* - x_i^*) +\\
                             &      & {r_i^d}^*(C_i^d - \nu_n^* - {x_i^d}^* - y_i^*+ x_i^*) +\\
                             &      & {\alpha_i^u}^*(-\sigma'_i({x_i^u}^*+y_i^*+x_i^*)) + \\
                             &      & {\alpha_i^d}^*(-\sigma'_i({x_i^d}^*+ y_i^*+x_i^*)) + \\
                             &       & \overline{P}_i (\rho_i^*+ x_i^*) + \overline{R}_i^u{x_i^u}^* + \overline{R}_i^d{x_i^d}^*.\\
\end{array}
\end{equation}
Using~\eqref{eq:dualcomp} in~\eqref{eq:genpart1}, it follows (recall that $i \in \I_n$, so $b(i) = n$) that 
\begin{equation}
\label{eq:genpart1b}
\Gamma_1^{g_i} \ge
-(\kappa_n^*- \sigma'_i{y^u_i}^*){\alpha_i^u}^* - {y^u_i}^*{r^u_i}^* - (\kappa_n^*- \sigma'_i{y^d_i}^*){\alpha_i^d}^* - {y^d_i}^*{r^d_i}^*.
\end{equation}
where the inequality follows from the fact that the terms in the last line of~\eqref{eq:genpart1} are non-negative; that is, $\overline{P}_i (\rho_i^*+ x_i^*) + \overline{R}_i^u{x_i^u}^* + \overline{R}_i^d{x_i^d}^* \ge 0$. Specifically, the parameters $\overline{P}_i,  \overline{R}_i^u, \overline{R}_i^d \ge 0$, and the optimal dual decision variables $\rho_i^*,x_i^*,{x_i^u}^*,{x_i^d}^*\ge 0$, as they are all associated to ``$\ge$'' constraints in the minimization LP~\eqref{eq:detCCmodel} (i.e., constraints~\eqref{eq:detuncer_con_limit}, \eqref{eq:detuncer_con2_limit}, \eqref{eq:detuncer_rampup_limit}, and~\eqref{eq:detuncer_rampdown_limit}, respectively).

Now, we claim (recall~\eqref{eq:tau}) that 
\begin{equation}
\label{eq:tauu}
\tau^u_i{r_i^u}^* \ge (\kappa_n^*- \sigma'_i{y^u_i}^*){\alpha_i^u}^* + {y^u_i}^*{r^u_i}^*.
\end{equation}
If $\kappa_{n}^*- \sigma'_i{y^u_i}^* \ge 0$, it follows from the lower bound constraint on $r_i^u$ in~\eqref{eq:detuncer_rampup_limit} that
$(\frac{\kappa_{n}^*}{\sigma'_i} - {y^u_i}^*) {r^u_i}^* \ge(\frac{\kappa_{n}^*}{\sigma'_i} - {y^u_i}^*)  \sigma'_i {\alpha^u_i}^*$; that is, $\frac{\kappa_{n}^*}{\sigma'_i}  {r^u_i}^* \ge(\kappa_{n}^*- \sigma'_i {y^u_i}^*){\alpha^u_i}^* +  {y^u_i}^*{r^u_i}^*$. If $\kappa_{n}^*- \sigma'_i{y^u_i}^* \le 0$, clearly 
${y^u_i}^*{r_i^u}^* \ge (\kappa_n^*- \sigma'_i{y^u_i}^*){\alpha_i^u}^* + {y^u_i}^*{r^u_i}^*$, since ${\alpha_i^u}^* \ge 0$. These two facts, together with~\eqref{eq:tau} show that~\eqref{eq:tauu} holds.

Similarly, using the lower bound constraint on $r_i^d$ in~\eqref{eq:detuncer_rampdown_limit}, it follows that
\begin{equation}
\label{eq:taud}
\tau^d_i{r_i^d}^* \ge (\kappa_n^*- \sigma'_i{y^d_i}^*){\alpha_i^d}^* + {y^d_i}^*{r^d_i}^*.
\end{equation}

Combining equations~\eqref{eq:genpart1b},~\eqref{eq:tauu}, and~\eqref{eq:taud} with the definition of $\Gamma^{g_i}_2$ in~\eqref{eq:detGENexprevenue}, it follows that  $\Gamma^{g_i} = \Gamma^{g_i}_1 + \Gamma^{g_i}_2 \ge 0$; that is,
the pricing scheme introduced in Definition~\ref{def:CCpricing} guarantees that in the CCO market-clearing model~\eqref{eq:CCmodel}, the expected revenue of  conventional generator $i$ at bus~$n$ in the market,  is greater than or equal than their expected operating costs (i.e., cost recovery is  guaranteed in expectation), for all $i \in \I_n, n \in \B$.

For the purpose of numerical analysis of the CC pricing scheme introduced in Definition~\ref{def:CCpricing}, it is useful to present the standard deviation of the conventional generators' profit $\sigma^{g_i}$; namely, from~\eqref{eq:winddef}, \eqref{eq:controlsrup}, \eqref{eq:controlsrdo}, and Definition~\ref{def:CCpricing}, it follows that for all $n \in \B$, $i \in \I_n$,
\begin{equation}
\label{eq:genstd}
\sigma^{g_i} = \left | {\alpha_i^u}^*(C_i^u - \nu_n^* - \tau_i^u) +  {\alpha_i^d}^*(C_i^d - \nu_n^* + \tau_i^d) \right | \sigma_n.
\end{equation}

\subsection{Consumers' Surplus}
For the purpose of numerical analysis of the CC pricing scheme introduced in Definition~\ref{def:CCpricing}, it is useful to present the expected value and the standard deviation of the consumers' surplus; namely (cf., Definition~\ref{def:CCpricing}), for all $j \in \J_n$, $n \in \B$, the expected value of the consumer's $j$ surplus is given by
\begin{equation}
\label{eq:custexpprofit}
\Gamma^{L_j} =  (\nu_n^* + \ccozeta)\mathbb{E}(\s_j^*)- (\lambda_n^* + \ccozeta)L_j
=  (\nu_n^* + \ccozeta)s_j^*- (\lambda_n^* + \ccozeta)L_j,
\end{equation}
and the standard deviation of the customer's $j$ surplus is given by
\begin{equation}
\label{eq:custvarprofit}
\sigma^{L_j} =  |\nu_n^* + \ccozeta|\gamma_j^*\sigma_n.
\end{equation}
Both the last equality in~\eqref{eq:custexpprofit} and~\eqref{eq:custvarprofit}  readily follow from 
\eqref{eq:winddef} and~\eqref{eq:controlscu}.

\section{Case Study}
\label{sec:casestudy}
\begin{figure}[!htb]
    \centering
    \includestandalone[scale = 0.45]{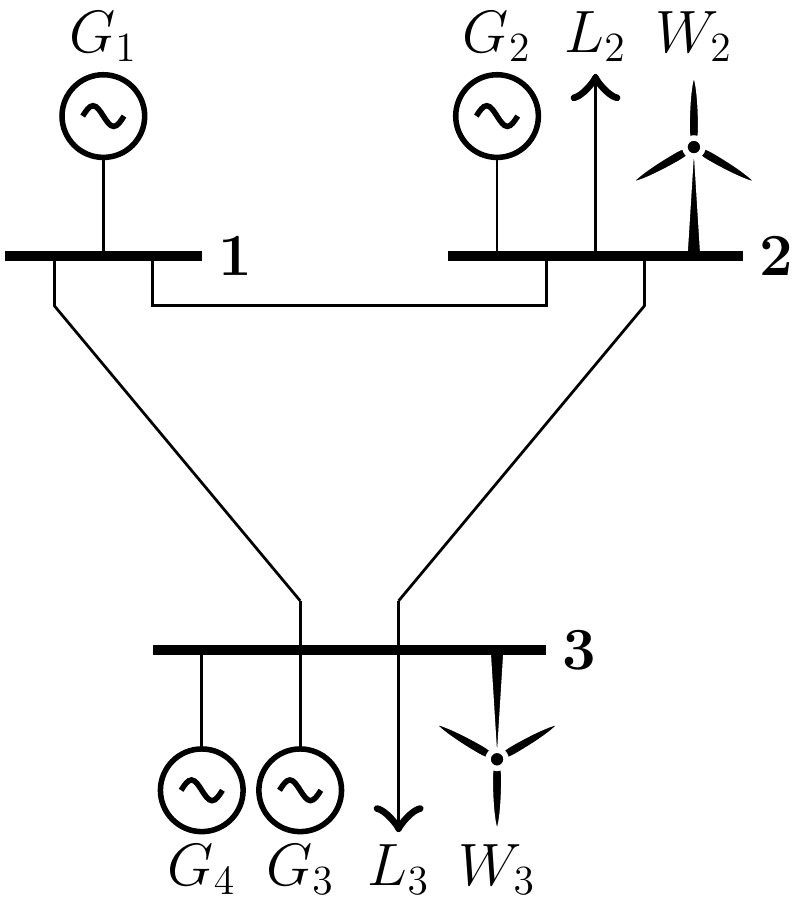}
    \caption{Network topology of case study.}
  \label{fig:Network}
\end{figure}

In order to illustrate the properties of the CCO market-clearing pricing scheme presented in Section~\ref{sec:CCPricing}, we next compute the prices and profits associated with the market participants of a simple power network whose topology is illustrated in Figure~\ref{fig:Network}.

Both the topology and the parameters of the power network in Figure~\ref{fig:Network}  are closely related to the case study network used in~\citet{morales2012pricing}; though some additional elements are added for discussion purposes. In particular, the following participants form the power network market: four (4) conventional generators ($G_1$, $G_2$, $G_3$, and $G_4$);  two (2) wind (i.e.,VRES) generators ($W_3$ and $W_2$), and two (2) loads ($L_2$ and $L_3$). Note that in labeling the network participants in the power network, we have taken the liberty to abuse notation and use labels for the wind generators and loads that are also used to define their associated parameters (cf., Table~\ref{tab:parnomenclature}).
The parameters associated with the power network in Figure~\ref{fig:Network} are provided in Table~\ref{tab:basicpars}, with the exception of the violation tolerance which is set to $\epsilon = 0.025$. To avoid repetition, in Table~\ref{tab:basicpars}, parameters that have equal values are separated by a comma; for example, for all conventional generators $i=1,\dots,4$, $C_i = C_i^u = C_i^d$.

\begin{table}[!htb]
\caption{Parameters associated with the case study network of Figure~\ref{fig:Network}.}
\label{tab:basicpars}
\centering
\footnotesize
\setlength{\tabcolsep}{3.5pt}
\begin{tabular}{lrrrrcclrrrcclrrcclrr}
\toprule
 & \multicolumn{4}{c}{Conv. Generators} & & & &\multicolumn{3}{c}{Lines}&& &&\multicolumn{2}{c}{Wind Gen.}& &&& \multicolumn{2}{c}{Loads}\\

 \cmidrule{2-5} \cmidrule{9-11} \cmidrule{15-16} \cmidrule{20-21}
 & \multicolumn{1}{c}{$G_1$} & \multicolumn{1}{c}{$G_2$} & \multicolumn{1}{c}{$G_3$} & \multicolumn{1}{c}{$G_4$} &&& & \multicolumn{1}{c}{$(1,2)$} & \multicolumn{1}{c}{$(1,3)$} & \multicolumn{1}{c}{$(2,3)$} &&&& \multicolumn{1}{c}{$W_2$} & \multicolumn{1}{c}{$W_3$} &&&& \multicolumn{1}{c}{$L_2$} & \multicolumn{1}{c}{$L_3$} \\
\midrule
$C, C^u, C^d$ & 20 & 25 & 30 & 22 &&& $B$ & $\frac{1}{0.13}$ & $\frac{1}{0.13}$ & $\frac{1}{0.13}$ &&&$C^w$ & 0 & 0 &&& $L$ & 70 & 200 \\
$\overline{P}$ & 100 & 50 & 100 & 20 &&& $\overline{C}$ & 100 & 60 & 100 &&& $\overline{W}, W^f$ & 80 & 34.50 &&& $V$ & 48.50 & 48.50\\
$\overline{R}^u, \overline{R}^d$ & 0 & 20 & 30 & 10 & & & & & & & & &$\sigma$ & 12 & 5.175 \\ 
\bottomrule
\end{tabular}
\end{table}

In Table~\ref{tab:basicpars}, we choose the standard deviation of the wind power forecasting errors to be 15\% of the day-ahead wind power forecast. This follows the statistical characteristics of data that we have analyzed. Namely, after looking at forecasted and actual values of wind power in the aggregated Belgian wind farms (\url{http://www.elia.be/en/grid-data/power-generation/wind-power}), one obtains that between 01/11/2018 and 01/11/2019, the standard deviation of the wind power forecasting error was $191.41$, whereas the average wind power forecasted in that period was $927.74$; that is. the standard deviation is about $20\%$ of the average wind power forecast.	

Note that the case study violates one of the assumptions made in Section~\ref{sec:prelim}; namely, $\sigma_1=0$ (i.e., there is no wind generator in bus~1). Loosely speaking, if bus $n' \in \B$ has no VRES generation, then one simply sets $W_{n'}^f= \overline{W}_{n'}= \sigma_{n'} = 0$, and $\alpha_i^u, \alpha_i^d, \beta_{n'}, \gamma_j = 0$, for all $i \in \I_{n'}, j \in \J_{n'}$, in~\eqref{eq:detCCmodel}. However, one has to be careful with the price definitions in~\eqref{eq:tau}. For a formal discussion on how to deal with buses without VRES generation, see Remark~\ref{rem:zero} in Appendix~\ref{sec:assumptions}.

After solving the CCO model~\eqref{eq:detCCmodel}, one obtains the results presented in Table~\ref{tab:case1sched},~\ref{tab:case1real}, and~\ref{tab:case1reve}. Table~\ref{tab:case1sched} presents the optimal values of the scheduling stage decision variables of the CCO model~\eqref{eq:detCCmodel} together with the corresponding prices obtained using Definition~\ref{def:CCpricing}. 

\begin{table}[!htb]
\caption{Optimal values and associated prices of market participant's actions in the scheduling stage of the electricity market in Figure~\ref{fig:Network}.}
\label{tab:case1sched}
\centering
\footnotesize
\begin{tabular}{ccccrrcr}
\toprule
Network& Action to &&&   &&  \\
Element&Compensate &&&Value  && Price\\
\midrule
$G_{1}$ & $p_{1}$ &&& 100.00  && 25.00 \\
$G_{2}$ & $p_{2}$ &&& 35.54  && 25.00 \\
$G_{3}$ & $p_{3}$ &&& 9.96  && 25.00 \\
$G_{4}$ & $p_{4}$ &&& 10.00  && 25.00 \\
$W_{2}$ & $w^{\rm sch}_{2}$ &&& 34.50  && 0.00 \\
$W_{3}$ & $w^{\rm sch}_{3}$ &&& 80.00  && 0.00 \\
$L_2$ & $s_{2}$ &&& 70.00  && 15.29 \\
$L_3$ & $s_{1}$ &&& 200.00  && 15.29  \\  
\bottomrule
\end{tabular}
\end{table}

Table~\ref{tab:case1real} presents the optimal values of the real-time stage decision variables of the CCO model~\eqref{eq:detCCmodel} (more precisely, e.g., the optimal value of $r_1^u = \mathbb{E}(\r_1^u)$ is given in the 3rd column of Table~\ref{tab:case1real}), together with the corresponding standard deviation (which can be readily computed from~\eqref{eq:controls} and~\eqref{eq:winddef}), and corresponding real-time prices obtained using Definition~\ref{def:CCpricing}. As Table~\ref{tab:case1real} illustrates, the latter prices are real-time uncertainty uniform; that is, they do not depend on the real-time realization of the VRES generation outcomes, which through~\eqref{eq:controls} affect the real-time values of upward and downward reserve, wind power spill, and load curtailment. This fact is further illustrated in Section~\ref{sec:vsstoc}. 

\begin{table}[htb]
\caption{Optimal values and associated prices of market participant's actions in the real-time stage of the electricity market in Figure~\ref{fig:Network}.}
\label{tab:case1real}
\centering
\footnotesize
\begin{tabular}{ccccrrcr}
\toprule
Network& Action to &&&  \multicolumn{2}{c}{Value} &&  \\
\cmidrule{5-6}
Element&Compensate &&& \multicolumn{1}{c}{Expected} & \multicolumn{1}{c}{Std. Dev.} && Price\\
\midrule
$G_{1}$ & $\r^u_{1}$ &&& 0.00 & (0.00) && 25.00 \\  
$G_{2}$ & $\r^u_{2}$ &&& 2.04 & (0.88) && 25.00  \\  
$G_{3}$ & $\r^u_{3}$ &&& 22.96 & (1.25) && 30.00 \\  
$G_{4}$ & $\r^u_{4}$ &&& 5.00 & (2.15) && 30.00  \\  
$G_{1}$ & $\r^d_{1}$ &&& 0.00 & (0.00) && 25.00 \\  
$G_{2}$ & $\r^d_{2}$ &&& 10.00 & (4.30) && 25.00 \\  
$G_{3}$ & $\r^d_{3}$ &&& 15.00 & (6.45) && 20.00  \\  
$G_{4}$ & $\r^d_{4}$ &&& 5.00 & (2.15) && 20.00 \\  
$W_{2}$ & $\w^{\rm spi}_{2}$ &&& 0.00 & (0.00) && 0.00 \\
$W_{3}$ & $\w^{\rm spi}_{3}$ &&& 0.00 & (0.00) && 0.00 \\  
$L_2$ & $\s_{2}$ &&& 0.00 & (0.00) && 15.29  \\
$L_3$ & $\s_{1}$ &&& 0.00 & (0.00) && 15.29  \\  
\bottomrule
\end{tabular}
\end{table}  

Table~\ref{tab:case1reve} presents the expected values and corresponding standard deviations of the electricity market participants' profit (cf., eq.~\eqref{eq:detISOexprevenue}, \eqref{eq:operstd}, \eqref{eq:detWINDexprevenue}, \eqref{eq:windstd},  \eqref{eq:detGENexprevenue}, \eqref{eq:genstd}, \eqref{eq:custexpprofit}, and~\eqref{eq:custvarprofit}). As proved in Theorem~\ref{thm:CCpricing}, the prices obtained in Tables~\ref{tab:case1sched} and~\ref{tab:case1real} ensure that, in the electricity market in Figure~\ref{fig:Network}, the market administrator has a non-negative expected profit, and that the revenue of the conventional  and the wind (i.e., VRES) generating units in the market (i.e., $G_1$, $G_2$, $G_3$, $G_4$, $W_3$, and $W_2$) is greater than or equal than their operating costs in expectation (cf., Table~\ref{tab:case1reve}). As shown in the 3rd column of Table~\ref{tab:case1reve}, the profits of the market participants vary depending on the realization of the wind's (i.e., VRES') power generation.

\begin{table}[htb]
\caption{Profits of market participants of the electricity market in Figure~\ref{fig:Network}.}
\label{tab:case1reve}
\centering
\footnotesize
\begin{tabular}{cccrr}
\toprule
    &&& \multicolumn{2}{c}{Profit} \\
    \cmidrule{4-5}
    &&& \multicolumn{1}{c}{Expected} & \multicolumn{1}{c}{Std. Dev.} \\
    \midrule
Market Administrator &&& 0.00 & (303.03) \\
$G_{1}$ &&& 500.00 & (0.00) \\
$G_{2}$ &&& 0.00 & (0.00) \\
$G_{3}$ &&& 100.22 & (64.48) \\
$G_{4}$ &&& 80.00 & (12.90) \\
$W_{2}$  &&& 0.00 & (0.00) \\
$W_{3}$  &&& 0.00 & (0.00) \\
$L_2$ &&& -1070.04 & (0.00) \\
$L_3$ &&& -3057.26 & (0.00) \\
\bottomrule
\end{tabular}
\end{table}  

All the numerical results presented in this section, as well as the ones presented in Section~\ref{sec:CCOvsSO} are obtained using a 1.7 GHz Intel Core i7 with a 8 GB 1600 MHz DDR3 RAM, using MATLAB R2019a together with YALMIP~\citep{lofberg2004yalmip} to formulate the optimization problems which are solved using CPLEX~12.8  with default parameters. The MATLAB code used to obtain the results of this section and those in Section~\ref{sec:CCOvsSO} is available upon request to the authors.

\section{A look at both Chance-Constrained and Stochastic prices.}
\label{sec:vsstoc}
In this section, we illustrate some differences and similarities between using the revenue adequate CCO pricing scheme studied here, and the revenue adequate SO pricing scheme introduced in~\citet{morales2012pricing}. For this purpose, and to make the article more self-contained, in Appendix~\ref{sec:stocmodel}, we present a  brief summary of this pricing scheme, using the same notation introduced in previous sections. Next, we analyze the results obtained from both pricing schemes.

\subsection{Illustrative comparison of CCO and SO pricing schemes.}
\label{sec:CCOvsSO}
The CCO market-clearing model~\eqref{eq:CCmodel}  and the SO market-clearing model~\eqref{eq:stocmodel} have a number of fundamental differences. In particular, the uncertainty associated with the VRES power generation is modeled differently. Namely, in the CCO model, the VRES power generation uncertainty is modeled by assuming that the continuous distribution of the power generation forecasting error is known (see, eq.~\eqref{eq:winddef}), whereas in the SO model, it is modeled by assuming that a finite discrete distribution of the VRES power generation is known (see, eq.~\eqref{eq:stocwinddef}). Also, the overall model uncertainty in the CCO model is handled through the use of chance constraints and real-time control rules~\citep[see, e.g.,][]{bienstock2014chance, lubin2015robust}, whereas in the SO model, it is modeled using a two-stage stochastic optimization model with recourse~\citep[see, e.g.,][]{morales2009economic, lamadrid2015}. 
Finally, while the SO pricing scheme uses dual information on the balance constraints in the scheduling and real-time stages from the SO model, the CCO pricing scheme uses additional dual information on the VRES power spill and generators' reserves (see Table~\ref{tab:prices} and~\eqref{eq:tau},~\eqref{def:kappa} for further details).
This means that results obtained from these models are not comparable in any rigorous way. Thus, the comparisons made in this section between the two pricing schemes only serve to illustrate some of their properties.

With this important points in mind, Definition~\ref{def:CCpricing}, which provides a revenue adequate pricing scheme for the CCO market-clearing model~\eqref{eq:CCmodel} (see Theorem~\ref{thm:CCpricing}), and Definition~\ref{def:stocpricing}~\citep{morales2012pricing}, which provides a revenue adequate pricing scheme for the SO market-clearing model~\eqref{eq:stocmodel} (see Theorem~\ref{thm:SOpricing}), can be used to illustratively compare the prices proposed by both models.

\subsubsection{Analytical comparison.}

Consider the analytical comparison between the CCO and SO pricing schemes in Table~\ref{tab:prices}. Note that in this table, we are, as mentioned in Section~\ref{sec:stocmodel}, abusing notation by using the same or similar labels for the dual variables associated to the balancing and re-balancing constraints in the CCO market-clearing model~\eqref{eq:CCmodel}, and the SO market-clearing model~\eqref{eq:stocmodel}. 
Also, as mentioned in Section~\ref{sec:casestudy}, note that in labeling the network participants in the power network (i.e., 2nd column in Table~\ref{tab:prices}), we have taken the liberty to abuse notation and use labels for the VRES generators and loads that are also used to define their associated parameters (cf., Table~\ref{tab:parnomenclature}).

\begin{table}[!htb]
\caption{Illustrative comparison of the chance-constrained and stochastic market-clearing, revenue adequate prices from Definition~\ref{def:CCpricing} and Definition~\ref{def:stocpricing}.}
\label{tab:prices}
\footnotesize
\centering
\begin{tabular}{lllccc}
\toprule
 & Network & &  & \multicolumn{2}{c}{Market-clearing model prices}\\
 \cmidrule{5-6}
 \multicolumn{1}{c}{Action at bus $n$} & \multicolumn{1}{c}{Element} &  \multicolumn{1}{c}{Stage} && Chance-Constrained & Stochastic \\
\midrule
Conventional Generation & $G_i$, $i \in \I_n$ & Scheduling &&  $\lambda_n$ & $\lambda_n$ \\
VRES Generation&     $W_n$    &Scheduling &&  $\lambda_n - y_n^{\spill} + x_n^{\spill}$ &$ \lambda_n$\\
Load Consumption & $L_j$, $j \in \J_n$ & Scheduling &&  $\lambda_n + \zeta$ & $\lambda_n$\\
\midrule
Load Curtailment & $L_j$, $j \in \J_n$ & Real-time && $\nu_n + \zeta$ & 
$\nu_n^{\omega}/\pi^{\omega}$, $\omega \in \Omega$\\
Upward Reserve&  $G_i$, $i \in \I_n$ & Real-time && $\nu_n+ \tau_i^u$ & 
$\nu_n^{\omega}/\pi^{\omega}$, $\omega \in \Omega$\\
Downward Reserve&  $G_i$, $i \in \I_n$ & Real-time && $\nu_n- \tau_i^d$ & 
$\nu_n^{\omega}/\pi^{\omega}$, $\omega \in \Omega$\\
VRES Power Surplus/Deficit & $W_n$   & Real-time && $\nu_n - y_n^{\spill} + x_n^{\spill}$ & $\nu_n^{\omega}/\pi^{\omega}$, $\omega \in \Omega$\\
\bottomrule
\end{tabular}
\end{table}

As the last four rows in Table~\ref{tab:prices} clearly show, the main difference between the CCO and the SO pricing schemes is that the CCO pricing scheme (cf., Definition~\ref{def:CCpricing}) provides prices for the real-time market participant's actions in the electricity market that do not depend on the real-time realization of the VRES' generation. On the other hand, Table~\ref{tab:prices} shows that both pricing schemes are similar, in the sense that the dual variables associated to the corresponding scheduling power balance constraints (labeled by $\lambda_n$ in both schemes) and real-time power rebalancing constraints (labeled by $\nu_n$ in the CCO model, and $\nu_n^{\omega}$ in the SO model) are key   in respectively defining the scheduling and real-time stage prices. However, in the SO pricing scheme, these dual values ($\lambda_n$, $\nu_n^{\omega}$) and the likelihood of the different VRES power generation scenarios ($\pi^{\omega}$) fully define the prices. This results in the SO prices being locationally uniform for all the market participants located in the same bus. In the CC pricing scheme, dual variables associated with the particular market participant affect the prices associated with upward and downward reserve and VRES generation surpluses/shortages (see the last three rows in Table~\ref{tab:prices}). 

Although to find both the SO and CC revenue adequate prices it is enough to solve a linear program, which current optimization solvers can solve very efficiently, even for fairly large problems, it is worth noting that the linear program that needs to be solved to obtain the CC prices is (in practice) much smaller that the one required to obtain the SO prices.

\begin{remark}
\label{rem:size}
To obtain:
\begin{enumerate}[label=(\roman*)]
\item the SO revenue-adequate prices in Definition~\ref{def:stocpricing}, one needs to solve the linear program~\eqref{eq:stocmodel} with $|\I|+2|\B|+|\Omega|(2|\I|+2|\B|+|\J|)$ variables and $|\B|+|\L|+|\I|+|\Omega|(2|\B|+|\L|+3|\I|+|\J|)$ constraints.
\item the CC revenue-adequate prices in Definition~\ref{def:CCpricing}, one needs to solve the linear program~\eqref{eq:detCCmodel} with $5|\I|+5|\B|+2|\J|$ variables and $6|I|+5|B|+2|J| +2|L|$ constraints.
\end{enumerate}
\end{remark}

Clearly, what defines the difference in size of the problems in Remark~\ref{rem:size} is the number of scenarios~$|\Omega|$ used in the SO market-clearing model~\eqref{eq:stocmodel}. Given that both the  SO and CC market-clearing models are linear programs, this difference is not very relevant, unless both $|\Omega|$ and the electricity network in consideration is very large. In particular, for the network in Figure~\ref{fig:Network}, while the CCO market-clearing model is solved in less than 1 sec, the SO market-clearing model is solved in about 8 sec. This difference however would become quite relevant if results like the ones presented here can be extended to market-clearing models in which commitment decisions are taken into account, as in that case, binary variables representing these decisions need to be added to the models. Solving such models in short times is greatly dependent on the size of the problem.

\subsubsection{Numerical comparison.}
\label{sec:numerical}
To further illustrate the discussion in the previous section, let us consider again the electricity market considered in Section~\ref{sec:casestudy} (i.e., defined by Figure~\ref{fig:Network} and Table~\ref{tab:basicpars}). To make the CCO and SO models somewhat consistent, in order to construct the SO market-clearing model~\eqref{eq:stocmodel}, we set the scenarios of the wind (i.e., VRES) power generation (cf.,~\eqref{eq:stocwinddef}), in buses $n \in \{2,3\}$, by setting the number of scenarios $|\Omega| = 1,000$, and independently sampling (for different buses) the values of $W^{\omega}_n$ for all $w \in \Omega$ from a normal distribution with mean $W_n^f$ and standard deviation $\sigma_n$, with the values of $W_n^f, \sigma_n$ given in Table~\ref{tab:basicpars}. Furthermore, the scenario probabilities are set by letting $\pi^{\omega} = |\Omega|^{-1}$ for all $\omega \in \Omega$. These numerical settings, together with the ones in Figure~\ref{fig:Network} and Table~\ref{tab:basicpars} define what we would refer as \case{1} in the discussion below. To numerically illustrate the differences between the  uncertainty uniform real-time prices in the CCO pricing scheme and the (scenario dependent) real-time  prices in the SO pricing schemes, we construct three additional cases based on \case{1} by purposely making modifications to some of the parameters of \case{1} so that the volatility of the SO real-time prices increases with respect to the SO real-time prices in \case{1} (shown in Table~\ref{tab:case1}). Specifically, \case{2} is a variation of \case{1} obtained by changing $\overline{R}^u, \overline{R}^d$ from $(0, 20, 30,10) \to (0, 10, 15,5)$ (cf., Table~\ref{tab:basicpars}). That is, \case{2} is obtained from \case{1} after a drastic $50\%$ reduction in the conventional generators' upward and downward reserve capacity, which reduces the capacity of the conventional generators to change their scheduled power output in the real-time stage. \case{3} is a variation of \case{1} obtained by changing $C^u$ from $(20, 25, 30,22) \to (21, 26.25, 31.5, 23.1)$ (cf., Table~\ref{tab:basicpars}), and changing $C^d$ from $(20, 25, 30,22) \to (19, 23.75, 28.5, 20.9)$. That is, \case{3} is obtained from \case{1} after increasing the upward reserve costs by $5\%$, and decreasing the downward reserve savings by $5\%$, which might lead to lower use of upward and downward reserve capabilities in the real-time stage to lower overall costs. \case{4} is a variation of \case{1} obtained by making both changes of parameters done in \case{2} and \case{3}.

\begin{table}[ht]
\caption{CCO and SO revenue adequate real-time prices for \case{1}. In the table,~$\mu$ indicates expectation, and~$\sigma$ indicates standard deviation.}
\label{tab:case1}
\footnotesize
\begin{minipage}[c]{0.45\linewidth}
\vspace{0pt}
\centering
\begin{tabular}{ccccrccrrccccccccccc}
\toprule
\multicolumn{2}{c}{} &&& \multicolumn{1}{c}{CCO} &&& \multicolumn{2}{c}{SO}\\
\multicolumn{2}{c}{Action} &&& \multicolumn{1}{c}{Prices}&&& \multicolumn{2}{c}{Prices}\\
\cmidrule{5-5} \cmidrule{8-9}
&    &&&&&&\multicolumn{1}{c}{$\mu$} & \multicolumn{1}{c}{$\sigma$}\\ 
\midrule
$G_{1}$ & $r^u_{1}$ &&&  25.00 &&& 25.57 & (2.01)\\  
$G_{2}$ & $r^u_{2}$ &&&  25.00 &&& 25.57 & (2.01)\\  
$G_{3}$ & $r^u_{3}$ &&&  30.00 &&& 25.57 & (2.01)\\  
$G_{4}$ & $r^u_{4}$ &&&  30.00 &&& 25.57 & (2.01)\\  
$G_{1}$ & $r^d_{1}$ &&&  25.00 &&& 25.57 & (2.01)\\  
$G_{2}$ & $r^d_{2}$ &&&  25.00 &&& 25.57 & (2.01)\\  
$G_{3}$ & $r^d_{3}$ &&&  20.00 &&& 25.57 & (2.01)\\  
$G_{4}$ & $r^d_{4}$ &&&  20.00 &&& 25.57 & (2.01)\\  
$W_{2}$ & $w^{\rm spi}_{2}$ &&&  0.00 &&& 25.57 & (2.01)\\  
$W_{3}$ & $w^{\rm spi}_{3}$ &&&  0.00 &&& 25.57 & (2.01)\\  
$L_2$ & $s_{2}$ &&&  15.29 &&& 25.57 & (2.01) \\
$L_3$ & $s_{1}$ &&&  15.29 &&& 25.57 & (2.01) \\
\bottomrule
\end{tabular}
\end{minipage}
\hspace{0.5cm}
\begin{minipage}[c]{0.45\linewidth}
\centering
\vspace{0pt}
\includegraphics[width=\textwidth]{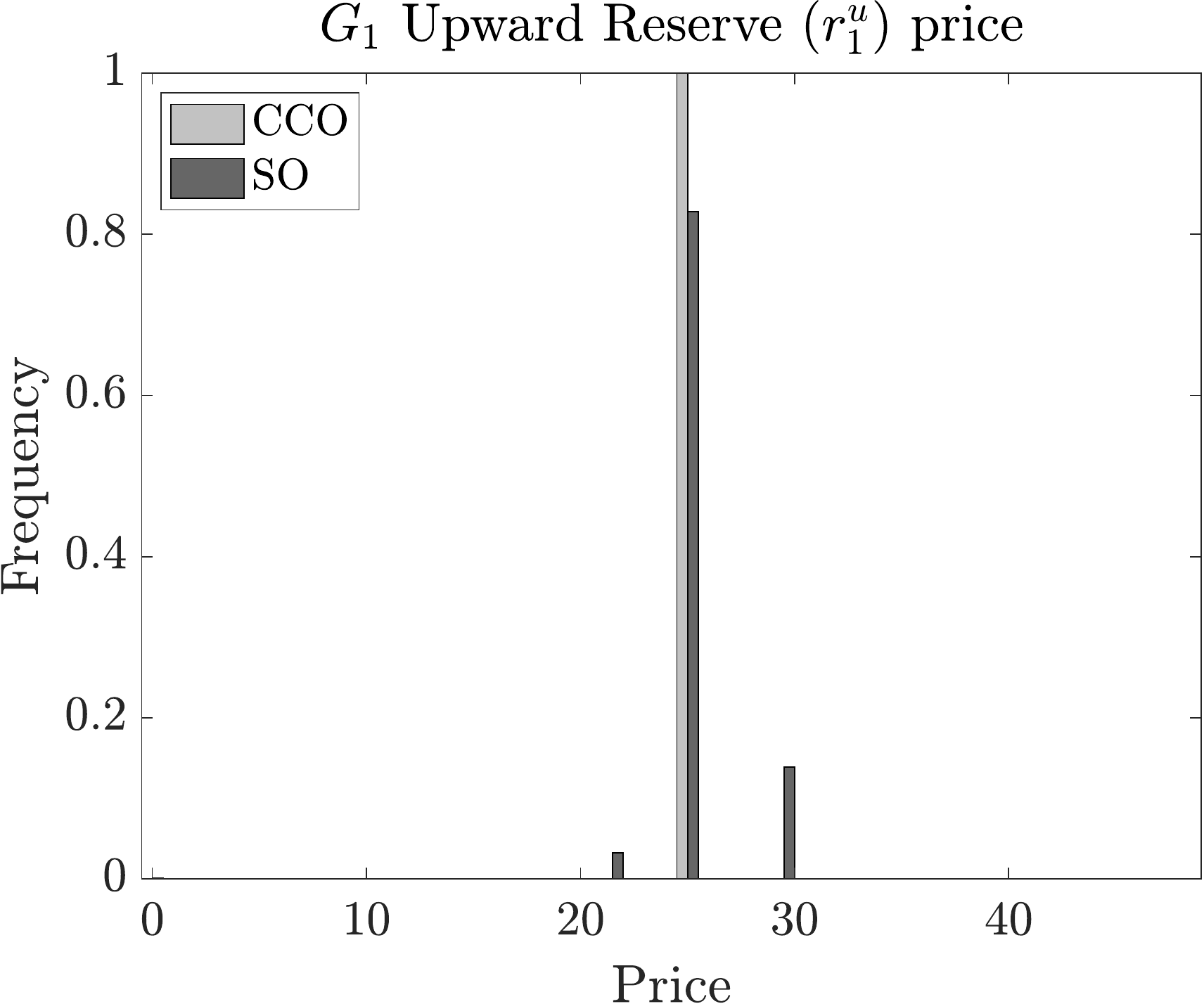}
\end{minipage}
\end{table}

In Table~\ref{tab:case1}, we present the real-time prices obtained using the revenue adequate CCO pricing scheme (Definition~\ref{def:CCpricing}, and Theorem~\ref{thm:CCpricing}) and the revenue adequate SO pricing scheme (Definition~\ref{def:stocpricing}, and Theorem~\ref{thm:SOpricing}). By real-time prices, we refer to the prices associated to all the actions taken by the participants in the real-time stage of the market-clearing process (i.e., upward and downward reserve deployments, wind power spillage, and load curtailment). As illustrated in the 3rd column of Table~\ref{tab:case1}, the real-time prices obtained from the CCO pricing scheme are uncertainty uniform, in the sense that they do not depend on the realizations of the uncertain wind power generated in the real-time stage. In contrast, as illustrated in the table's 4th and 5th columns, the real-time prices obtained from the SO pricing scheme vary depending on the realizations of the uncertain wind real-time power generated. In particular, these columns respectively show the expected value (indicated by $\mu$) and the standard deviation (indicated by $\sigma$) of these prices. To further illustrate this point, in Table~\ref{tab:case1}, we plot the distribution associated with the price of conventional generator $G_1$'s upward reserve for both the CCO and SO pricing schemes. This plot shows that depending on the realizations of the uncertain wind real-time power generated the SO price can take three different values between 20 and 30, whereas the CCO price is always the same. Note that the CCO price of 25.00 is close but not equal to the expected value of the SO price of 25.57 (cf., first row, 3rd and 4th column in Table~\ref{tab:case1}). One might think that this proximity in value is to be intuitively expected. However, as Table~\ref{tab:case1} shows, the CCO prices can be quite different to the expected value of the corresponding SO prices.


\begin{table}[ht]
\caption{CCO and SO revenue adequate real-time prices for \case{2}. In the table,~$\mu$ indicates expectation, and~$\sigma$ indicates standard deviation.}
\label{tab:case2}
\footnotesize
\begin{minipage}[c]{0.45\linewidth}
\vspace{0pt}
\centering
\begin{tabular}{ccccrccrrccccccccccc}
\toprule
\multicolumn{2}{c}{} &&& \multicolumn{1}{c}{CCO} &&& \multicolumn{2}{c}{SO}\\
\multicolumn{2}{c}{Action} &&& \multicolumn{1}{c}{Prices}&&& \multicolumn{2}{c}{Prices}\\
\cmidrule{5-5} \cmidrule{8-9}
&    &&&&&&\multicolumn{1}{c}{$\mu$} & \multicolumn{1}{c}{$\sigma$}\\ 
\midrule
$G_{1}$ & $r^u_{1}$ &&&  24.25 &&& 25.00 & (6.32)\\
$G_{2}$ & $r^u_{2}$ &&&  48.50 &&& 25.00 & (6.32) \
$G_{3}$ & $r^u_{3}$ &&&  48.50 &&& 25.00 & (6.32)\\
$G_{4}$ & $r^u_{4}$ &&&  48.50 &&& 25.00 & (6.32)\\
$G_{1}$ & $r^d_{1}$ &&&  24.25 &&& 25.00 & (6.32)\\
$G_{2}$ & $r^d_{2}$ &&&  0.00 &&& 25.00 & (6.32) \\
$G_{3}$ & $r^d_{3}$ &&&  0.00 &&& 25.00 & (6.32)\\
$G_{4}$ & $r^d_{4}$ &&&  0.00 &&& 25.00 & (6.32)\\
$W_{2}$ & $w^{\rm spi}_{2}$ &&&  0.00 &&& 25.00& (6.32)  \\
$W_{3}$ & $w^{\rm spi}_{3}$ &&&  0.00 &&& 25.00 & (6.32) \\
$L_2$ & $s_{2}$ &&&  16.97 &&& 25.00 & (6.32)\\
$L_3$ & $s_{1}$ &&&  16.97 &&& 25.00 & (6.32) \\
\bottomrule
\end{tabular}
\end{minipage}
\hspace{0.5cm}
\begin{minipage}[c]{0.45\linewidth}
\centering
\vspace{0pt}
\includegraphics[width=\textwidth]{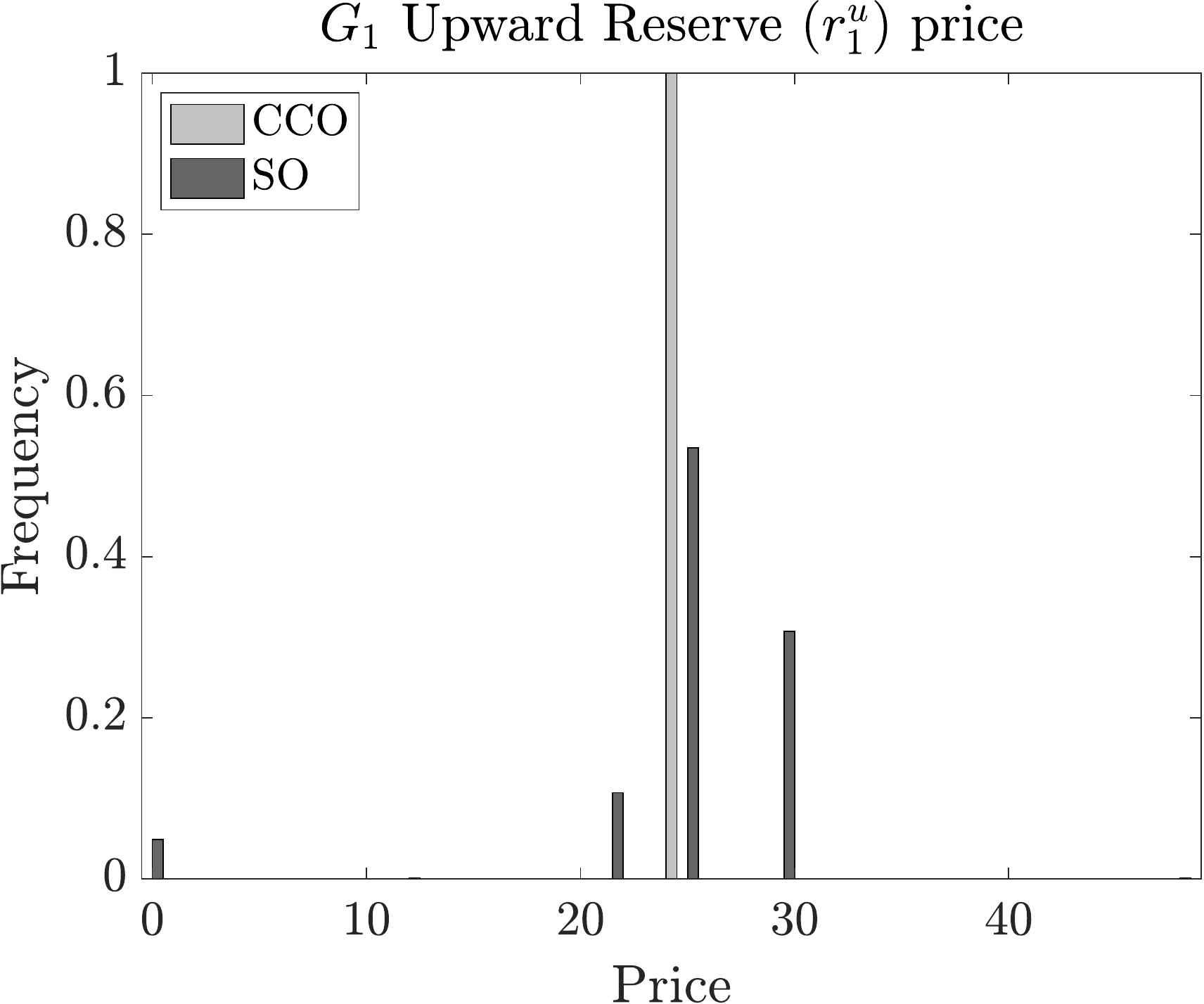}
\end{minipage}
\end{table}

Table~\ref{tab:case1} also illustrate the fact that the SO prices are locationally uniform; that is, have the same value for all the market participants' actions in the same bus, whereas the CCO prices might not be equal for market participants located in the same bus. For example, the prices associated with the upward reserve by $G_2$, the downward reserve by $G_2$, the wind power spill by $W_2$, and the curtailment of $L_2$, that are all located in bus~2 (cf., Figure~\ref{fig:Network}), are different.

Table~\ref{tab:case2}, Table~\ref{tab:case3}, and Table~\ref{tab:case4}, illustrate the same points discussed above for \case{1}, but this time for \case{2}, \case{3}, and \case{4}, respectively. As one goes through these results, the main difference is that the cases are purposely set up to show instances of the market associated to Figure~\ref{fig:Network} in which the volatility of the corresponding SO real-time prices increases (in comparison to \case{1}). This again highlights that whereas CCO real-time prices are independent of the realizations of the uncertain wind real-time power generated, the SO real-time prices can vary significantly. In \case{1}, the volatility (i.e., standard deviation) of the real-time prices is  
about 8\% (of the prices' expected value), in \case{2} is about 25\%, in \case{3} is about 19\%, and in \case{4} is about 33\%, and the number of different prices associated with the realizations of uncertain wind power generation goes from 3 to 7.


\begin{table}[ht]
\caption{CCO and SO revenue adequate real-time prices for \case{3}. In the table,~$\mu$ indicates expectation, and~$\sigma$ indicates standard deviation.}
\label{tab:case3}
\footnotesize
\begin{minipage}[c]{0.45\linewidth}
\vspace{0pt}
\centering
\begin{tabular}{ccccrccrrccccccccccc}
\toprule
\multicolumn{2}{c}{} &&& \multicolumn{1}{c}{CCO} &&& \multicolumn{2}{c}{SO}\\
\multicolumn{2}{c}{Action} &&& \multicolumn{1}{c}{Prices}&&& \multicolumn{2}{c}{Prices}\\
\cmidrule{5-5} \cmidrule{8-9}
&    &&&&&&\multicolumn{1}{c}{$\mu$} & \multicolumn{1}{c}{$\sigma$}\\ 
\midrule
$G_{1}$ & $r^u_{1}$ &&&  25.00 &&& 24.95 & (4.79) \\
$G_{2}$ & $r^u_{2}$ &&&  26.25 &&& 24.95 & (4.79) \\
$G_{3}$ & $r^u_{3}$ &&&  31.50 &&& 24.95 & (4.79) \\
$G_{4}$ & $r^u_{4}$ &&&  31.50 &&& 24.95 & (4.79) \\
$G_{1}$ & $r^d_{1}$ &&&  25.00 &&& 24.95 & (4.79) \\
$G_{2}$ & $r^d_{2}$ &&&  23.75 &&& 24.95 & (4.79) \\
$G_{3}$ & $r^d_{3}$ &&&  18.50 &&& 24.95 & (4.79) \\
$G_{4}$ & $r^d_{4}$ &&&  18.50 &&& 24.95 & (4.79) \\
$W_{2}$ & $w^{\rm spi}_{2}$ &&&  0.00 &&& 24.95 & (4.79) \\
$W_{3}$ & $w^{\rm spi}_{3}$ &&&  0.00 &&& 24.95 & (4.79) \\
$L_2$ & $s_{2}$ &&&  15.13 &&& 24.95 & (4.79) \\
$L_3$ & $s_{1}$ &&&  15.13 &&& 24.95 & (4.79) \\
\bottomrule
\end{tabular}
\end{minipage}
\hspace{0.5cm}
\begin{minipage}[c]{0.45\linewidth}
\centering
\vspace{0pt}
\includegraphics[width=\textwidth]{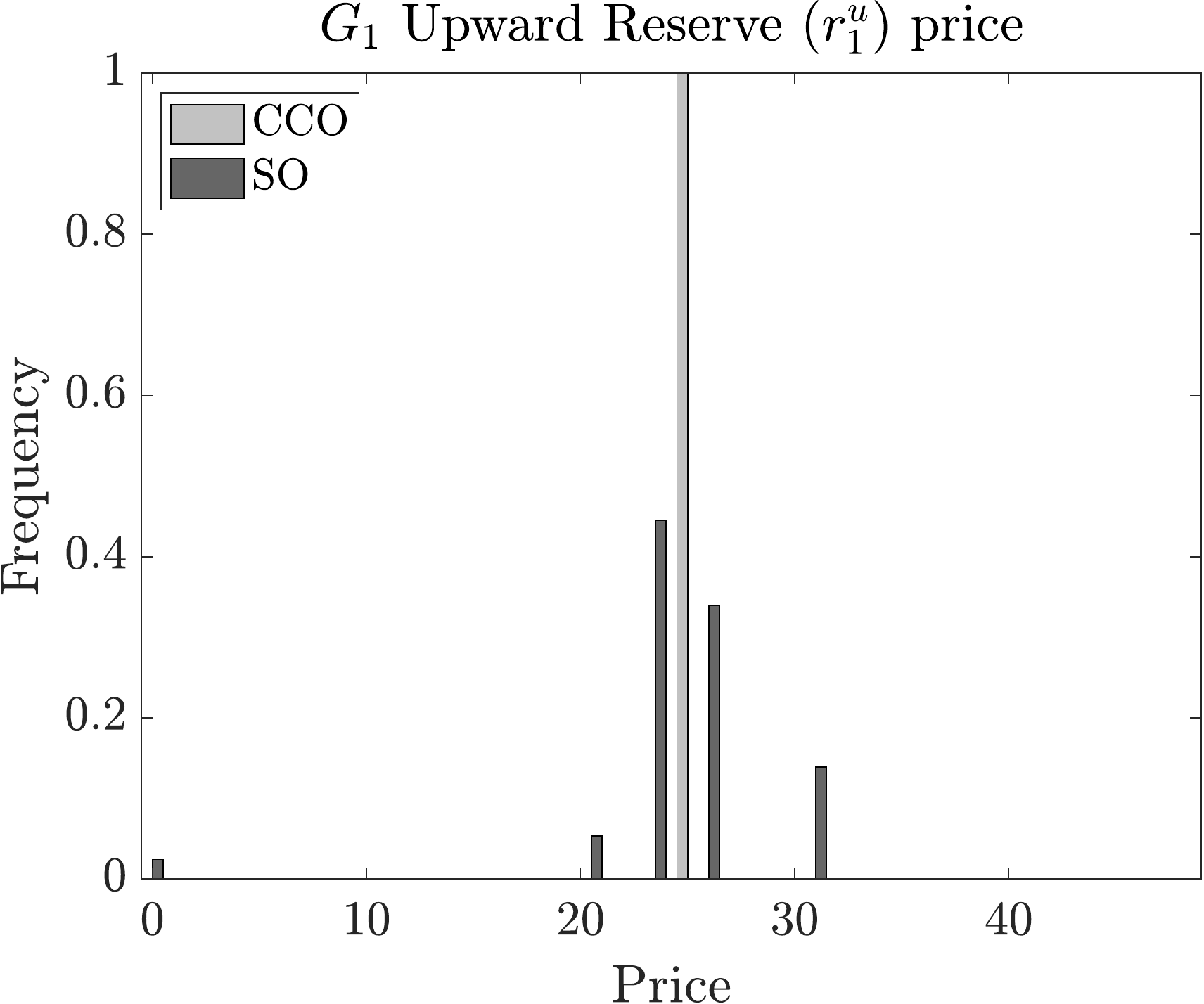}
\end{minipage} 
\end{table}

Note that the plot in Table~\ref{tab:case4} shows that the SO price of upward reserve for conventional generator $G_1$ could be equal to about 48. The value is precisely 48.50, which is equal to the market administrator's cost for load curtailment (cf., Table~\ref{tab:basicpars}). This curtailment cost might be considered too low. However, if the value of this cost was set to 1000~\citep[as in][]{morales2012pricing}, the histogram in Table~\ref{tab:case4} would be exactly the same, except that the bar at value 48.50  would be located at value 1,000. This is the reason why the cost of curtailment was chosen to be ``just'' twice the average cost of conventional power generation in the market. Otherwise, a value like 1,000 would make the graphical presentation of the results difficult, but more importantly, this ``outlier'' price would render the corresponding values of the mean price as well as their standard deviation presented in Tables~\ref{tab:case1}-\ref{tab:case4}, meaningless.

\begin{table}[ht]
\caption{CCO and SO revenue adequate real-time prices for \case{4}. In the table,~$\mu$ indicates expectation, and~$\sigma$ indicates standard deviation.}
\label{tab:case4}
\footnotesize
\begin{minipage}[c]{0.45\linewidth}
\vspace{0pt}
\centering
\begin{tabular}{ccccrccrrccccccccccc}
\toprule
\multicolumn{2}{c}{} &&& \multicolumn{1}{c}{CCO} &&& \multicolumn{2}{c}{SO}\\
\multicolumn{2}{c}{Action} &&& \multicolumn{1}{c}{Prices}&&& \multicolumn{2}{c}{Prices}\\
\cmidrule{5-5} \cmidrule{8-9}
&    &&&&&&\multicolumn{1}{c}{$\mu$} & \multicolumn{1}{c}{$\sigma$}\\ 
\midrule
$G_{1}$ & $r^u_{1}$ &&&  24.25 &&& 24.13 & (7.98) \\
$G_{2}$ & $r^u_{2}$ &&&  48.50 &&& 24.13 & (7.98) \\
$G_{3}$ & $r^u_{3}$ &&&  48.50 &&& 24.13 & (7.98) \\
$G_{4}$ & $r^u_{4}$ &&&  48.50 &&& 24.13 & (7.98) \\
$G_{1}$ & $r^d_{1}$ &&&  24.25 &&& 24.13 & (7.98) \\
$G_{2}$ & $r^d_{2}$ &&&  0.00 &&& 24.13 & (7.98) \\
$G_{3}$ & $r^d_{3}$ &&&  0.00 &&& 24.13 & (7.98) \\
$G_{4}$ & $r^d_{4}$ &&&  0.00 &&& 24.13 & (7.98) \\
$W_{2}$ & $w^{\rm spi}_{2}$ &&&  0.00 &&& 24.13& (7.98)  \\
$W_{3}$ & $w^{\rm spi}_{3}$ &&&  0.00 &&& 24.13 & (7.98) \\
$L_2$ & $s_{2}$ &&&  16.97 &&& 24.13 & (7.98) \\
$L_3$ & $s_{1}$ &&&  16.97 &&& 24.13 & (7.98) \\
\bottomrule
\end{tabular}
\end{minipage}
\hspace{0.5cm}
\begin{minipage}[c]{0.45\linewidth}
\centering
\vspace{0pt}
\includegraphics[width=\textwidth]{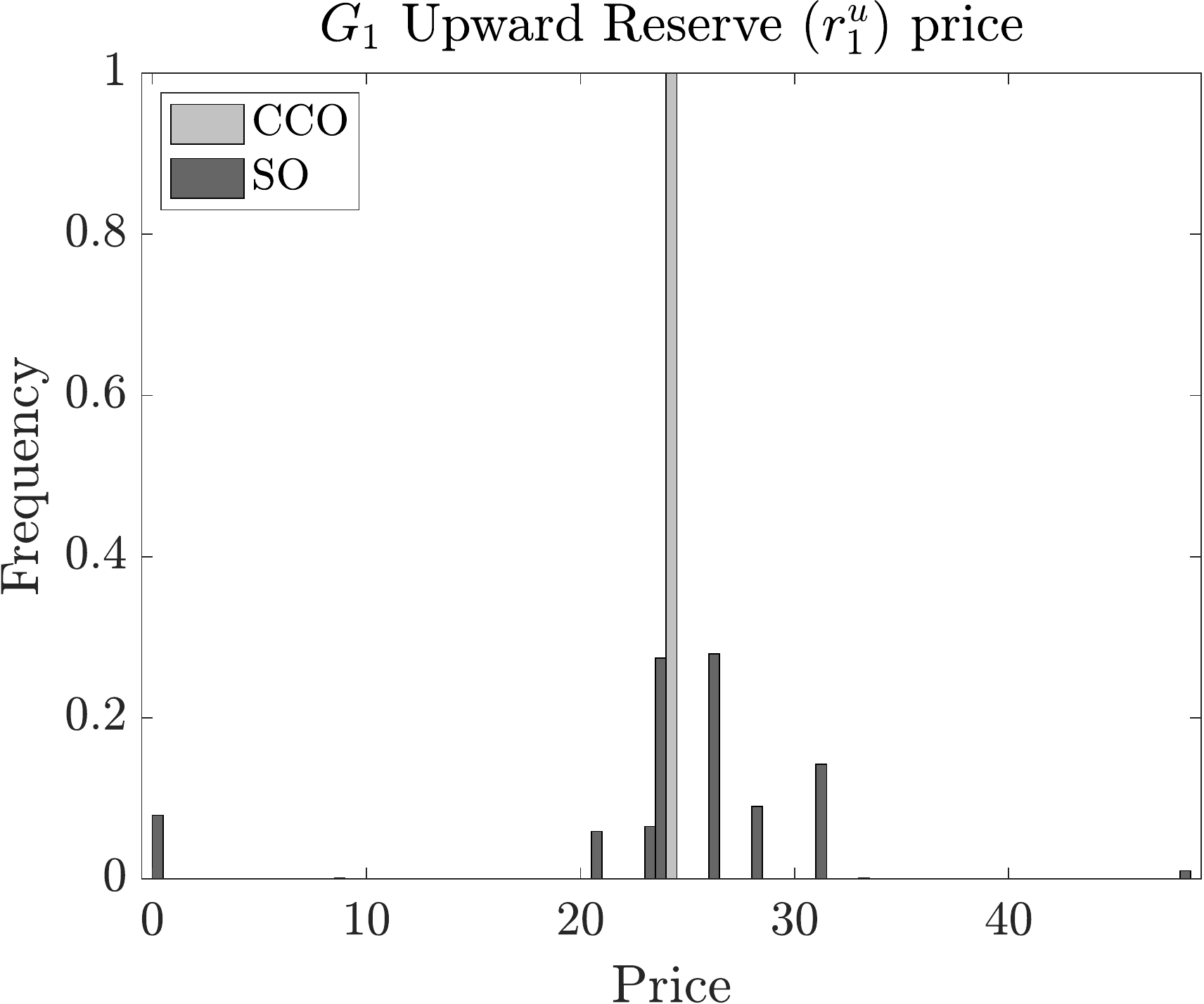}
\end{minipage}
\end{table}

Table~\ref{tab:revenues12} and Table~\ref{tab:revenues34} show the expected profits and standard deviation of the profits for market participants in both the CCO and SO pricing scheme for \case{1}, \case{2}, \case{3}, and \case{4}.
As mentioned in Section~\ref{sec:casestudy}, these values are calculated using~\eqref{eq:detISOexprevenue}, \eqref{eq:operstd}, \eqref{eq:detWINDexprevenue}, \eqref{eq:windstd},  \eqref{eq:detGENexprevenue}, \eqref{eq:genstd}, \eqref{eq:custexpprofit}, and~\eqref{eq:custvarprofit}, for the CCO pricing scheme. The corresponding values for the SO pricing scheme are calculated using \eqref{eq:stocISOexprevenue},
\eqref{eq:stocISOstd},
\eqref{eq:stocWINDexprevenue},
\eqref{eq:stocWINDstd},
\eqref{eq:stocGENexprevenue},
\eqref{eq:stocGENstd},
\eqref{eq:stocCUSTexprevenue}, and
\eqref{eq:stocCUSTstd}, in Appendix~\ref{sec:stocrevs}.
Beyond illustrating that, as designed, the prices defined  by both the CCO and SO pricing scheme result in the expected profit of the market administrator, conventional generators, and wind generators to be revenue adequate (i.e., non-negative), the results hint that, when compared to SO revenues, the CCO pricing scheme leads to higher (less negative) expected profits for the loads at the ``cost'' of leading to an increased volatility of the market administrator's expected profit. Also, the results hint that, when compared to SO revenues, the CCO pricing scheme leads to lower overall volatility in the expected profit of conventional generators, and substantially less volatility in the expected profit of wind generators. However, more evidence and study that will be the focus of future research, is required to conclude whether these are general differences between the CCO and SO pricing schemes in terms of expected profits and corresponding volatilities.

\begin{table}[ht]
\caption{Expected profits ($\mu$) and standard deviation of profits ($\sigma$) of the market participants in the CCO and SO revenue adequate pricing schemes for~\case{1} and~\case{2}.}
\label{tab:revenues12}
\footnotesize
\begin{minipage}[c]{0.45\linewidth}
\vspace{0pt}
\centering
\footnotesize
\setlength{\tabcolsep}{2.5pt}
\begin{tabular}{cccrrccrrccccccccccc}
\toprule
\multicolumn{1}{c}{} &&& \multicolumn{2}{c}{\case{1}} &&& \multicolumn{2}{c}{\case{1}}\\
\multicolumn{1}{c}{Actor} &&& \multicolumn{2}{c}{CCO Profits}&&& \multicolumn{2}{c}{SO Profits}\\
\cmidrule{4-5} \cmidrule{8-9}
&    &&\multicolumn{1}{c}{$\mu$} & \multicolumn{1}{c}{$\sigma$}&&&\multicolumn{1}{c}{$\mu$} & \multicolumn{1}{c}{$\sigma$}\\ 
\midrule
Oper. &&& 0.00 & (303.03) &&& 0.00 & (0.00) \\ 
$G_{1}$ &&& 500.00 & (0.00) &&& 557.40 & (0.00) \\ 
$G_{2}$ &&& 0.00 & (0.00) &&& 33.54 & (38.59) \\ 
$G_{3}$ &&& 100.22 & (64.48) &&& 0.00 & (31.23) \\ 
$G_{4}$ &&& 80.00 & (12.90) &&& 71.92 & (19.34) \\ 
$W_{2}$  &&& 0.00 & (0.00) &&& 884.38 & (132.68) \\ 
$W_{3}$  &&& 0.00 & (0.00) &&& 2041.85 & (316.19) \\ 
$L_2$ &&& -1070.04 & (0.00) &&& -1790.18 & (0.00) \\ 
$L_3$ &&& -3057.26 & (0.00) &&& -5114.80 & (0.00) \\ 
\bottomrule
\end{tabular}
\end{minipage}
\hspace{0.5cm}
\begin{minipage}[c]{0.45\linewidth}
\vspace{0pt}
\centering
\footnotesize
\setlength{\tabcolsep}{2.5pt}
\begin{tabular}{cccrrccrrccccccccccc}
\toprule
\multicolumn{1}{c}{} &&& \multicolumn{2}{c}{\case{2}} &&& \multicolumn{2}{c}{\case{2}}\\
\multicolumn{1}{c}{Actor} &&& \multicolumn{2}{c}{CCO Profits}&&& \multicolumn{2}{c}{SO Profits}\\
\cmidrule{4-5} \cmidrule{8-9}
&    &&\multicolumn{1}{c}{$\mu$} & \multicolumn{1}{c}{$\sigma$}&&&\multicolumn{1}{c}{$\mu$} & \multicolumn{1}{c}{$\sigma$}\\ 
\midrule
Oper. &&& 0.00 & (266.07) &&& 0.00 & (0.00) \\ 
$G_{1}$ &&& 500.00 & (0.00) &&& 500.00 & (0.00) \\ 
$G_{2}$ &&& 242.50 & (3.22) &&& 31.17 & (55.01) \\ 
$G_{3}$ &&& 288.75 & (37.08) &&& 0.55 & (94.39) \\ 
$G_{4}$ &&& 166.25 & (4.84) &&& 70.87 & (23.55) \\ 
$W_{2}$  &&& 25.88 & (0.00) &&& 856.35 & (130.09) \\ 
$W_{3}$  &&& 60.00 & (0.00) &&& 1957.04 & (290.75) \\ 
$L_2$ &&& -1156.16 & (36.32) &&& -1750.00 & (0.00) \\ 
$L_3$ &&& -3544.71 & (0.00) &&& -4999.94 & (1.97) \\ 
\bottomrule
\end{tabular}
\end{minipage}
\end{table}

\begin{table}[ht]
\caption{Expected profits ($\mu$) and standard deviation of profits ($\sigma$) of the market participants in the CCO and SO revenue adequate pricing schemes for~\case{3} and~\case{4}.}
\label{tab:revenues34}
\footnotesize
\begin{minipage}[c]{0.45\linewidth}
\vspace{0pt}
\centering
\footnotesize
\setlength{\tabcolsep}{2.5pt}
\begin{tabular}{cccrrccrrccccccccccc}
\toprule
\multicolumn{1}{c}{} &&& \multicolumn{2}{c}{\case{3}} &&& \multicolumn{2}{c}{\case{3}}\\
\multicolumn{1}{c}{Actor} &&& \multicolumn{2}{c}{CCO Profits}&&& \multicolumn{2}{c}{SO Profits}\\
\cmidrule{4-5} \cmidrule{8-9}
&    &&\multicolumn{1}{c}{$\mu$} & \multicolumn{1}{c}{$\sigma$}&&&\multicolumn{1}{c}{$\mu$} & \multicolumn{1}{c}{$\sigma$}\\ 
\midrule
Oper. &&& 0.00 & (345.67) &&& -0.00 & (0.00) \\ 
$G_{1}$ &&& 500.00 & (0.00) &&& 572.98 & (0.00) \\ 
$G_{2}$ &&& 0.00 & (0.00) &&& 50.91 & (75.94) \\ 
$G_{3}$ &&& 0.00 & (21.40) &&& 0.00 & (0.00) \\ 
$G_{4}$ &&& 84.00 & (12.90) &&& 79.61 & (31.99) \\ 
$W_{2}$  &&& 0.00 & (0.00) &&& 883.63 & (129.32) \\ 
$W_{3}$  &&& 0.00 & (0.00) &&& 2027.82 & (301.42) \\ 
$L_2$ &&& -1058.82 & (0.00) &&& -1801.08 & (0.00) \\ 
$L_3$ &&& -3025.19 & (0.00) &&& -5145.95 & (0.00) \\ 
\bottomrule
\end{tabular}
\end{minipage}
\hspace{0.5cm}
\begin{minipage}[c]{0.45\linewidth}
\vspace{0pt}
\centering
\footnotesize
\setlength{\tabcolsep}{2.5pt}
\begin{tabular}{cccrrccrrccccccccccc}
\toprule
\multicolumn{1}{c}{} &&& \multicolumn{2}{c}{\case{4}} &&& \multicolumn{2}{c}{\case{4}}\\
\multicolumn{1}{c}{Actor} &&& \multicolumn{2}{c}{CCO Profits}&&& \multicolumn{2}{c}{SO Profits}\\
\cmidrule{4-5} \cmidrule{8-9}
&    &&\multicolumn{1}{c}{$\mu$} & \multicolumn{1}{c}{$\sigma$}&&&\multicolumn{1}{c}{$\mu$} & \multicolumn{1}{c}{$\sigma$}\\ 
\midrule
Oper. &&& 0.00 & (266.07) &&& 0.00 & (0.00) \\ 
$G_{1}$ &&& 500.00 & (0.00) &&& 500.00 & (0.00) \\ 
$G_{2}$ &&& 230.00 & (3.22) &&& 32.79 & (66.27) \\ 
$G_{3}$ &&& 266.25 & (37.08) &&& 2.58 & (42.15) \\ 
$G_{4}$ &&& 160.75 & (4.84) &&& 68.32 & (28.09) \\ 
$W_{2}$  &&& 25.88 & (0.00) &&& 852.59 & (129.93) \\ 
$W_{3}$  &&& 60.00 & (0.00) &&& 1935.60 & (297.60) \\ 
$L_2$ &&& -1156.16 & (36.32) &&& -1749.80 & (6.29) \\
$L_3$ &&& -3544.71 & (0.00) &&& -4998.21 & (23.38) \\ 
\bottomrule
\end{tabular}
\end{minipage}
\end{table}

\section{Final Remarks}
\label{sec:final}
In this article, we derive a market-clearing model~\eqref{eq:detCCmodel}
 with a corresponding revenue adequate pricing scheme (cf., Definition~\ref{def:CCpricing} and Theorem~\ref{thm:CCpricing}) in which there is uncertainty in the power generated by variable renewable energy sources (VRES). In contrast with the more popular choice of using scenarios to model the inherent uncertainty in VRES generation, in the proposed market-clearing model, we use distributional information about the deviations of the VRES power generation from its forecasted value and chance-constrained optimization techniques. These results show that 
revenue adequate prices in expectation can be obtained for an electricity dispatch model in which uncertainties are handled using chance-constrained optimization techniques. 
A fundamental difference between revenue adequate, stochastic market-clearing prices and revenue adequate, chance-constrained market-clearing prices is that the former prices  are dependent on the real-time outcome of the uncertainties in the market for actions taken in the real-time stage. In contrast, the latter real-time prices are uncertainty uniform in the sense that they do not depend on the real-time uncertainty outcomes. This fact is illustrated analytically and numerically in Sections~\ref{sec:casestudy} and~\ref{sec:vsstoc}.   
 
 \looseness=-1
Dispatch models using chance-constrained optimization~\citep[see, e.g.,][]{nemirovski2006convex} techniques to model market uncertainties have become very relevant in the recent literature~\citep[see, e.g.,][]{pozo2012chance, 
 bienstock2014chance, lubin2015robust, zhang2017chance, halilbavsic2018convex,
lubin2019chance, venzke2020chance}. To our knowledge, the chance-constrained market-clearing pricing scheme presented here is the first one to provide prices that ensure revenue adequacy in expectation for the market administrator, and cost recovery in expectation for both the conventional and VRES generators participating in the market, in a chance-constrained optimization dispatch model. In turn, this sets the basis to obtain similar pricing schemes for other chance-constrained optimization dispatch models. In particular, notice that a key difference between the dispatch model considered here~(cf.,~\eqref{eq:CCmodel}) and the dispatch model considered in~\citet{bienstock2014chance} is the fact that the affine controls are written in terms of a linear combination of the uncertain quantities as opposed to a single uncertain quantity in~\eqref{eq:controls}. This together with an independence assumption allows to reformulate the dispatch problem as a second-order cone optimization (SOCO) problem~\citep[see, e.g.,][]{lobo1998second}. Thus, in light of the results presented here, it is natural to consider extending the pricing scheme in Definition~\ref{def:CCpricing} to a dispatch model like the one presented in~\citet{bienstock2014chance} using SOCO duality beyond the LP duality used to prove Theorem~\ref{thm:CCpricing}.

To the best of our knowledge, there are no results in the literature looking at obtaining revenue adequate prices for uncertain marking-clearing models (e.g., taking into account VRES generation uncertainty) in which unit commitment decisions are considered (recall the literature reviewed in the article's Introduction). On the other hand, a few articles have considered the problem of obtaining equilibrium prices for uncertain marking-clearing models in which commitment decisions are considered~\citep[see, e.g.,][]{wang2011wind, mazadi2013impact, kuang2018pricing, kuang2019pricing}. We believe that the results presented here pave the way to extend them to obtain revenue adequate prices for uncertain marking-clearing models in which commitment decisions are considered.

Finally, it is worth mentioning that similar to, e.g., \citet{bienstock2014chance}, the chance-constrained formulation of the dispatch model is obtained by restricting the chance of violation of individual real-time constraints. As discussed in~\citet[][Section 1.4]{bienstock2014chance} and in further detail in~\citet{nemirovski2006convex}, more refined formulations restricting the chance of violation of a group of (or all) the real-time constraints could be studied. However, as discussed in~\citet{nemirovski2006convex}, obtaining a ``tractable'' deterministic reformulations or good tractable approximations for such problems is difficult, even under strong assumptions. On the other hand, advances in this direction have been recently obtained in, for example,~\citet{lubin2015two} to handle two-sided constraints (e.g., constraints setting upper and lower bounds on a real-time decision variable). Exploring the derivation of revenue adequate prices for more refined chance-constrained formulations of the market will be a line of future work.

\bibliographystyle{apalike}
\bibliography{CC_Pricing_BibTeX}

\begin{thebibliography}{}

\bibitem[Abbaspourtorbati et~al., 2016]{abbaspourtorbati2016pricing}
Abbaspourtorbati, F., Conejo, A.~J., Wang, J., and Cherkaoui, R. (2016).
\newblock Pricing electricity through a stochastic non-convex market-clearing
  model.
\newblock {\em IEEE Transactions on Power Systems}, 32(2):1248--1259.

\bibitem[Ben-Tal et~al., 2009]{ben2009robust}
Ben-Tal, A., El~Ghaoui, L., and Nemirovski, A. (2009).
\newblock {\em Robust optimization}, volume~28.
\newblock Princeton University Press.

\bibitem[Bienstock et~al., 2014]{bienstock2014chance}
Bienstock, D., Chertkov, M., and Harnett, S. (2014).
\newblock Chance-constrained optimal power flow: Risk-aware network control
  under uncertainty.
\newblock {\em SIAM Review}, 56(3):461--495.

\bibitem[Birge and Louveaux, 2011]{birge2011introduction}
Birge, J.~R. and Louveaux, F. (2011).
\newblock {\em Introduction to stochastic programming}.
\newblock Springer Science \& Business Media.

\bibitem[Bjorndal et~al., 2016]{bjorndal2016congestion}
Bjorndal, E., Bjorndal, M.~H., Midthun, K., and Zakeri, G. (2016).
\newblock Congestion management in a stochastic dispatch model for electricity
  markets.
\newblock {\em NHH Dept. of Business and Management Science Discussion Paper},
  (2016/12).
\newblock Available at
  \url{https://openaccess.nhh.no/nhh-xmlui/bitstream/handle/11250/2401597/1216.pdf?sequence=1&isAllowed=y}.

\bibitem[Bj{\o}rndal and J{\"o}rnsten, 2008]{bjorndal2008equilibrium}
Bj{\o}rndal, M. and J{\"o}rnsten, K. (2008).
\newblock Equilibrium prices supported by dual price functions in markets with
  non-convexities.
\newblock {\em European Journal of Operational Research}, 190(3):768--789.

\bibitem[Bose, 2015]{bose2015design}
Bose, S. (2015).
\newblock On the design of wholesale electricity markets under uncertainty.
\newblock In {\em 2015 53rd Annual Allerton Conference on Communication,
  Control, and Computing (Allerton)}, pages 203--210. IEEE.

\bibitem[Bouffard et~al., 2005]{bouffard2005market}
Bouffard, F., Galiana, F.~D., and Conejo, A.~J. (2005).
\newblock Market-clearing with stochastic security-part {I}: formulation.
\newblock {\em IEEE Transactions on Power Systems}, 20(4):1818--1826.

\bibitem[Boyd and Vandenberghe, 2004]{boyd2004convex}
Boyd, S. and Vandenberghe, L. (2004).
\newblock {\em Convex optimization}.
\newblock Cambridge university press.

\bibitem[Chen et~al., 2018]{chen2018distributionally}
Chen, Y., Guo, Q., Sun, H., Li, Z., Wu, W., and Li, Z. (2018).
\newblock A distributionally robust optimization model for unit commitment
  based on kullback--leibler divergence.
\newblock {\em IEEE Transactions on Power Systems}, 33(5):5147--5160.

\bibitem[Chvatal et~al., 1983]{chvatal1983linear}
Chvatal, V., Chvatal, V., et~al. (1983).
\newblock {\em Linear programming}.
\newblock Macmillan.

\bibitem[Ding et~al., 2016]{ding2016robust}
Ding, T., Wu, Z., Lv, J., Bie, Z., and Zhang, X. (2016).
\newblock Robust co-optimization to energy and ancillary service joint dispatch
  considering wind power uncertainties in real-time electricity markets.
\newblock {\em IEEE Transactions on Sustainable Energy}, 7(4):1547--1557.

\bibitem[Dvorkin, 2019]{dvorkin2019chance}
Dvorkin, Y. (2019).
\newblock A chance-constrained stochastic electricity market.
\newblock {\em arXiv preprint arXiv:1906.06963}.

\bibitem[Dvorkin et~al., 2015]{dvorkin2015uncertainty}
Dvorkin, Y., Lubin, M., Backhaus, S., and Chertkov, M. (2015).
\newblock Uncertainty sets for wind power generation.
\newblock {\em IEEE Transactions on Power Systems}, 31(4):3326--3327.

\bibitem[Fox and Bajari, 2013]{fox2013}
Fox, J.~T. and Bajari, P. (2013).
\newblock Measuring the efficiency of an fcc spectrum auction.
\newblock {\em American economic journal: Microeconomics}, 5(1):100--146.

\bibitem[Halilba{\v{s}}i{\'c} et~al., 2018]{halilbavsic2018convex}
Halilba{\v{s}}i{\'c}, L., Pinson, P., and Chatzivasileiadis, S. (2018).
\newblock Convex relaxations and approximations of chance-constrained ac-opf
  problems.
\newblock {\em IEEE Transactions on Power Systems}, 34(2):1459--1470.

\bibitem[Hodge and Milligan, 2011]{hodge2011wind}
Hodge, B.-M. and Milligan, M. (2011).
\newblock Wind power forecasting error distributions over multiple timescales.
\newblock In {\em 2011 IEEE power and energy society general meeting}, pages
  1--8. IEEE.

\bibitem[Hortacsu and Puller, 2008]{hortacsu2008}
Hortacsu, A. and Puller, S.~L. (2008).
\newblock Understanding strategic bidding in multi-unit auctions: a case study
  of the texas electricity spot market.
\newblock {\em The RAND Journal of Economics}, 39(1):86--114.

\bibitem[Jabr, 2013]{jabr2013adjustable}
Jabr, R.~A. (2013).
\newblock Adjustable robust opf with renewable energy sources.
\newblock {\em IEEE Transactions on Power Systems}, 28(4):4742--4751.

\bibitem[Jayantilal et~al., 2001]{jayantilal2001market}
Jayantilal, A., Cheung, K.~W., Shamsollahi, P., and Bresler, F.~S. (2001).
\newblock Market based regulation for the {PJM} electricity market.
\newblock In {\em PICA 2001. Innovative Computing for Power-Electric Energy
  Meets the Market. 22nd IEEE Power Engineering Society. International
  Conference on Power Industry Computer Applications (Cat. No. 01CH37195)},
  pages 155--160. IEEE.

\bibitem[Joskow et~al., 1998]{joskow1998}
Joskow, P.~L., Schmalensee, R., and Bailey, E.~M. (1998).
\newblock The market for sulfur dioxide emissions.
\newblock {\em The American Economic Review}, 88(4):669--685.

\bibitem[Kazempour et~al., 2018]{kazempour2018stochastic}
Kazempour, J., Pinson, P., and Hobbs, B.~F. (2018).
\newblock A stochastic market design with revenue adequacy and cost recovery by
  scenario: Benefits and costs.
\newblock {\em IEEE Transactions on Power Systems}, 33(4):3531--3545.

\bibitem[Khazaei et~al., 2013]{khazaei2013market}
Khazaei, J., Zakeri, G., and Oren, S. (2013).
\newblock Market clearing mechanisms under demand uncertainty.
\newblock {\em Tech. rep., Electric Power Optimization Centre, University of
  Auckland}.

\bibitem[Khazaei et~al., 2017]{khazaei2017single}
Khazaei, J., Zakeri, G., and Oren, S.~S. (2017).
\newblock Single and multisettlement approaches to market clearing under demand
  uncertainty.
\newblock {\em Operations Research}, 65(5):1147--1164.

\bibitem[Kramer et~al., 2018]{kramer2018strictly}
Kramer, A., Krebs, V., and Schmidt, M. (2018).
\newblock Strictly and $\gamma$-robust counterparts of electricity market
  models: Perfect competition and nash-cournot equilibria.
\newblock Technical report, Optimization Online.
\newblock Available at
  \url{http://www.optimization-online.org/DB_HTML/2018/07/6709.html}.

\bibitem[Kuang et~al., 2018]{kuang2018pricing}
Kuang, X., Dvorkin, Y., Lamadrid, A.~J., Ortega-Vazquez, M.~A., and Zuluaga,
  L.~F. (2018).
\newblock Pricing chance constraints in electricity markets.
\newblock {\em IEEE Transactions on Power Systems}, 33(4):4634--4636.

\bibitem[Kuang et~al., 2019]{kuang2019pricing}
Kuang, X., Lamadrid, A.~J., and Zuluaga, L.~F. (2019).
\newblock Pricing in non-convex markets with quadratic deliverability costs.
\newblock {\em Energy Economics}, 80:123--131.

\bibitem[Lamadrid et~al., 2015]{lamadrid2015}
Lamadrid, A., Shawhan, D., Murillo-Sanchez, C., Zimmerman, R., Zhu, Y.,
  Tylavsky, D., Kindle, A., and Dar, Z. (2015).
\newblock Stochastically optimized, carbon-reducing dispatch of storage,
  generation, and loads.
\newblock {\em Power Systems, IEEE Transactions on}, 30(2):1064 -- 1075.

\bibitem[{Lamadrid} et~al., 2019]{lamadrid2019}
{Lamadrid}, A.~J., {Mu{\~n}oz-{\'A}lvarez}, D., {Murillo-S{\'a}nchez}, C.~E.,
  {Zimmerman}, R.~D., {Shin}, H., and {Thomas}, R.~J. (2019).
\newblock Using the {Matpower Optimal Scheduling Tool} to test power system
  operation methodologies under uncertainty.
\newblock {\em IEEE Transactions on Sustainable Energy}, 10(3):1280--1289.

\bibitem[Lange, 2005]{lange2005uncertainty}
Lange, M. (2005).
\newblock On the uncertainty of wind power predictions: {A}nalysis of the
  forecast accuracy and statistical distribution of errors.
\newblock {\em Journal of Solar Energy Engeneering}, 127(2):177--184.

\bibitem[Lawton et~al., 2003]{lawton2003}
Lawton, L., Sullivan, M., Van~Liere, K., Katz, A., and Eto, J. (2003).
\newblock A framework and review of customer outage costs: Integration and
  analysis of electric utility outage cost surveys.
\newblock Technical report, U.S. Department of Energy, Washington DC.

\bibitem[Liberopoulos and Andrianesis, 2016]{liberopoulos2016critical}
Liberopoulos, G. and Andrianesis, P. (2016).
\newblock Critical review of pricing schemes in markets with non-convex costs.
\newblock {\em Operations Research}, 64(1):17--31.

\bibitem[Lobo et~al., 1998]{lobo1998second}
Lobo, M.~S., Vandenberghe, L., Boyd, S., and Lebret, H. (1998).
\newblock Second-order cone programming.
\newblock {\em Linear algebra and Applications}, 284:193--228.

\bibitem[L{\"o}fberg, 2004]{lofberg2004yalmip}
L{\"o}fberg, J. (2004).
\newblock Yalmip: A toolbox for modeling and optimization in {MATLAB}.
\newblock In {\em Proceedings of the CACSD Conference}, volume~3. Taipei,
  Taiwan.

\bibitem[Lorca et~al., 2016]{lorca2016multistage}
Lorca, {\'A}., Sun, X.~A., Litvinov, E., and Zheng, T. (2016).
\newblock Multistage adaptive robust optimization for the unit commitment
  problem.
\newblock {\em Operations Research}, 64(1):32--51.

\bibitem[Lubin et~al., 2015a]{lubin2015two}
Lubin, M., Bienstock, D., and Vielma, J.~P. (2015a).
\newblock Two-sided linear chance constraints and extensions.
\newblock {\em arXiv preprint arXiv:1507.01995}.

\bibitem[Lubin et~al., 2015b]{lubin2015robust}
Lubin, M., Dvorkin, Y., and Backhaus, S. (2015b).
\newblock A robust approach to chance constrained optimal power flow with
  renewable generation.
\newblock {\em IEEE Transactions on Power Systems}, 31(5):3840--3849.

\bibitem[Lubin et~al., 2019]{lubin2019chance}
Lubin, M., Dvorkin, Y., and Roald, L. (2019).
\newblock Chance constraints for improving the security of ac optimal power
  flow.
\newblock {\em IEEE Transactions on Power Systems}, 34(3):1908--1917.

\bibitem[Mazadi et~al., 2013]{mazadi2013impact}
Mazadi, M., Rosehart, W., Zareipour, H., Malik, O., and Oloomi, M. (2013).
\newblock Impact of wind integration on electricity markets: a
  chance-constrained nash cournot model.
\newblock {\em International Transactions on Electrical Energy Systems},
  23(1):83--96.

\bibitem[Milgrom, 2000]{milgrom2000}
Milgrom, P. (2000).
\newblock Putting auction theory to work: The simultaneous ascending auction.
\newblock {\em Journal of Political Economy}, 108(2):245--272.

\bibitem[Morales et~al., 2012]{morales2012pricing}
Morales, J.~M., Conejo, A.~J., Liu, K., and Zhong, J. (2012).
\newblock Pricing electricity in pools with wind producers.
\newblock {\em IEEE Transactions on Power Systems}, 27(3):1366--1376.

\bibitem[Morales et~al., 2013]{morales2013integrating}
Morales, J.~M., Conejo, A.~J., Madsen, H., Pinson, P., and Zugno, M. (2013).
\newblock {\em Integrating renewables in electricity markets: operational
  problems}, volume 205.
\newblock Springer Science \& Business Media.

\bibitem[Morales et~al., 2009]{morales2009economic}
Morales, J.~M., Conejo, A.~J., and P{\'e}rez-Ruiz, J. (2009).
\newblock Economic valuation of reserves in power systems with high penetration
  of wind power.
\newblock {\em IEEE Transactions on Power Systems}, 24(2):900--910.

\bibitem[Nemirovski and Shapiro, 2006]{nemirovski2006convex}
Nemirovski, A. and Shapiro, A. (2006).
\newblock Convex approximations of chance constrained programs.
\newblock {\em SIAM Journal on Optimization}, 17(4):969--996.

\bibitem[O'Neill et~al., 2005]{ref.oneill2005}
O'Neill, R.~P., Sotkiewicz, P.~M., Hobbs, B.~F., Rothkopf, M.~H., and Stewart,
  W.~R. (2005).
\newblock Efficient market-clearing prices in markets with nonconvexities.
\newblock {\em European Journal of Operational Research}, 164(1):269--285.

\bibitem[Ozturk et~al., 2004]{ozturk2004solution}
Ozturk, U.~A., Mazumdar, M., and Norman, B.~A. (2004).
\newblock A solution to the stochastic unit commitment problem using chance
  constrained programming.
\newblock {\em IEEE Transactions on Power Systems}, 19(3):1589--1598.

\bibitem[Pereira and Pinto, 1991]{pereira1991multi}
Pereira, M.~V. and Pinto, L.~M. (1991).
\newblock Multi-stage stochastic optimization applied to energy planning.
\newblock {\em Mathematical Programming}, 52(1-3):359--375.

\bibitem[Pozo and Contreras, 2012]{pozo2012chance}
Pozo, D. and Contreras, J. (2012).
\newblock A chance-constrained unit commitment with an $n-k$ security criterion
  and significant wind generation.
\newblock {\em IEEE Transactions on Power systems}, 28(3):2842--2851.

\bibitem[Pritchard et~al., 2010]{pritchard2010single}
Pritchard, G., Zakeri, G., and Philpott, A. (2010).
\newblock A single-settlement, energy-only electric power market for
  unpredictable and intermittent participants.
\newblock {\em Operations research}, 58(4-part-2):1210--1219.

\bibitem[Roald et~al., 2016]{roald2016corrective}
Roald, L., Misra, S., Krause, T., and Andersson, G. (2016).
\newblock Corrective control to handle forecast uncertainty: A chance
  constrained optimal power flow.
\newblock {\em IEEE Transactions on Power Systems}, 32(2):1626--1637.

\bibitem[Roald et~al., 2015]{roald2015security}
Roald, L., Oldewurtel, F., Van~Parys, B., and Andersson, G. (2015).
\newblock Security constrained optimal power flow with distributionally robust
  chance constraints.
\newblock {\em arXiv preprint arXiv:1508.06061}.

\bibitem[Ruiz et~al., 2012]{ref.ruiz2012pricing}
Ruiz, C., Conejo, A.~J., and Gabriel, S.~A. (2012).
\newblock Pricing non-convexities in an electricity pool.
\newblock {\em IEEE Transactions on Power Systems}, 27(3):1334--1342.

\bibitem[Scarf, 1990]{scarf1990mathematical}
Scarf, H.~E. (1990).
\newblock Mathematical programming and economic theory.
\newblock {\em Operations Research}, 38(3):377--385.

\bibitem[Schweppe et~al., 1988]{schweppe1988}
Schweppe, F., Caramanis, M., Tabors, R., and Bohn, R. (1988).
\newblock {\em Spot Pricing of Electricity}.
\newblock The Kluwer International Series in Engineering and Computer Science :
  Power electronics and power systems. Springer.

\bibitem[Sen et~al., 2006]{sen2006stochastic}
Sen, S., Yu, L., and Genc, T. (2006).
\newblock A stochastic programming approach to power portfolio optimization.
\newblock {\em Operations Research}, 54(1):55--72.

\bibitem[Venzke et~al., 2020]{venzke2020chance}
Venzke, A., Halilba{\v{s}}i{\'c}, L., Barr{\'e}, A., Roald, L., and
  Chatzivasileiadis, S. (2020).
\newblock Chance-constrained ac optimal power flow integrating hvdc lines and
  controllability.
\newblock {\em International Journal of Electrical Power \& Energy Systems},
  116:105522.

\bibitem[Wang et~al., 2011]{wang2011wind}
Wang, Q., Wang, J., and Guan, Y. (2011).
\newblock Wind power bidding based on chance-constrained optimization.
\newblock In {\em 2011 IEEE Power and Energy Society General Meeting}, pages
  1--2. IEEE.

\bibitem[Wong and Fuller, 2007]{wong2007pricing}
Wong, S. and Fuller, J.~D. (2007).
\newblock Pricing energy and reserves using stochastic optimization in an
  alternative electricity market.
\newblock {\em IEEE Transactions on Power Systems}, 22(2):631--638.

\bibitem[Woo et~al., 1991]{woo1991}
Woo, C.~K., Pupp, R.~L., Flaim, T., and Mango, R. (1991).
\newblock How much do electric customers want to pay for reliability; new
  evidence on an old controversy.
\newblock {\em Energy Systems and Policy}, 15:2:Pages: 145--159.
\newblock ID: 4434610247.

\bibitem[Xu et~al., 2018]{xu2018optimal}
Xu, B., Shi, Y., Kirschen, D.~S., and Zhang, B. (2018).
\newblock Optimal battery participation in frequency regulation markets.
\newblock {\em IEEE Transactions on Power Systems}, 33(6):6715--6725.

\bibitem[Ye et~al., 2016]{ye2016uncertainty}
Ye, H., Ge, Y., Shahidehpour, M., and Li, Z. (2016).
\newblock Uncertainty marginal price, transmission reserve, and day-ahead
  market clearing with robust unit commitment.
\newblock {\em IEEE Transactions on Power Systems}, 32(3):1782--1795.

\bibitem[Zakeri et~al., 2018]{zakeri2018pricing}
Zakeri, G., Pritchard, G., Bjorndal, M., and Bjorndal, E. (2018).
\newblock Pricing wind: A revenue adequate, cost recovering uniform price
  auction for electricity markets with intermittent generation.
\newblock {\em INFORMS Journal on Optimization}, 1(1):35--48.

\bibitem[Zavala et~al., 2017]{zavala2017stochastic}
Zavala, V.~M., Kim, K., Anitescu, M., and Birge, J. (2017).
\newblock A stochastic electricity market clearing formulation with consistent
  pricing properties.
\newblock {\em Operations Research}, 65(3):557--576.

\bibitem[Zhang et~al., 2017]{zhang2017chance}
Zhang, Y., Wang, J., Zeng, B., and Hu, Z. (2017).
\newblock Chance-constrained two-stage unit commitment under uncertain load and
  wind power output using bilinear benders decomposition.
\newblock {\em IEEE Transactions on Power Systems}, 32(5):3637--3647.

\bibitem[Zhou et~al., 2016]{zhou2016stochastic}
Zhou, Z., Liu, C., and Botterud, A. (2016).
\newblock Stochastic methods applied to power system operations with renewable
  energy: A review.
\newblock Technical report, Argonne National Lab.(ANL), Argonne, IL (United
  States).
\newblock Available at
  \url{https://publications.anl.gov/anlpubs/2016/08/129468.pdf}.

\bibitem[Zou et~al., 2018]{zou2018partially}
Zou, J., Ahmed, S., and Sun, X.~A. (2018).
\newblock Partially adaptive stochastic optimization for electric power
  generation expansion planning.
\newblock {\em INFORMS Journal on Computing}, 30(2):388--401.

\end{thebibliography}

\newpage

\begin{appendices}

\section{About Assumptions}
\label{sec:assumptions}
In this section, we discuss in further detail some of the assumptions made in Section~\ref{sec:CCmodel} for the derivation of the  DCCO market-clearing model~\eqref{eq:detCCmodel} and the associated pricing scheme~(cf., Definition~\ref{def:CCpricing}). 

We assume in~\eqref{eq:winddef} that $\Delta \W_n^f$ is normally distributed for all $n \in \B$. However, all the results stated in Section~\ref{sec:CCmodel} and Section~\ref{sec:CCPricing} hold as long as $\Delta \W_n^f$ for all $n \in \B$ are random variables with zero mean and a symmetric (univariate) continuous distribution. Let $\mathbb{Q}(\widetilde{\Delta \W}_n^f)^{1-\epsilon}$ be the $1-\epsilon$ quantile of $\widetilde{\Delta \W}_n^f := (\Delta \W_n^f- \mathbb{E}(\Delta \W_n^f))/\sigma(\Delta \W_n^f)$ for any $0 \le \epsilon \le 1$.

\begin{remark}
\label{rem:dist}
Let $n' \in \B$ be a node in which $\Delta \W_n$ is not normally distributed but follows a symmetric univariate continuous distribution with zero mean. Then, the DCCO model~\eqref{eq:detCCmodel} is correct, and Theorem~\ref{thm:CCpricing} holds, after replacing  $\Phi^{-1}_{1-\epsilon}$ by $\mathbb{Q}(\widetilde{\Delta \W}_n^f)_{1-\epsilon}$  in  all constraints in~\eqref{eq:detCCmodel} involving  bus $n'$.
\end{remark}

This remark follows as the deterministic equivalents of the chance constraints discussed before stating the DCCO market-clearing model~\eqref{eq:detCCmodel} hold for any continuous distribution, as long as the normal distribution's $1-\epsilon$ quantile $\Phi^{-1}_{1-\epsilon}$ is replaced by the $\mathbb{Q}(\widetilde{\Delta \W}_n^f)_{1-\epsilon}$.
For example if $\Delta \W_n^f \sim \operatorname{Uniform}[-a,a]$ for some $a >0$, then one should change $\Phi^{-1}_{1-\epsilon} \to \sqrt{3}(1-2\epsilon)$ in all constraints in~\eqref{eq:detCCmodel}  involving  bus $n'$. 

\begin{figure}[!htb]
    \centering
    \includegraphics[scale = 0.45]{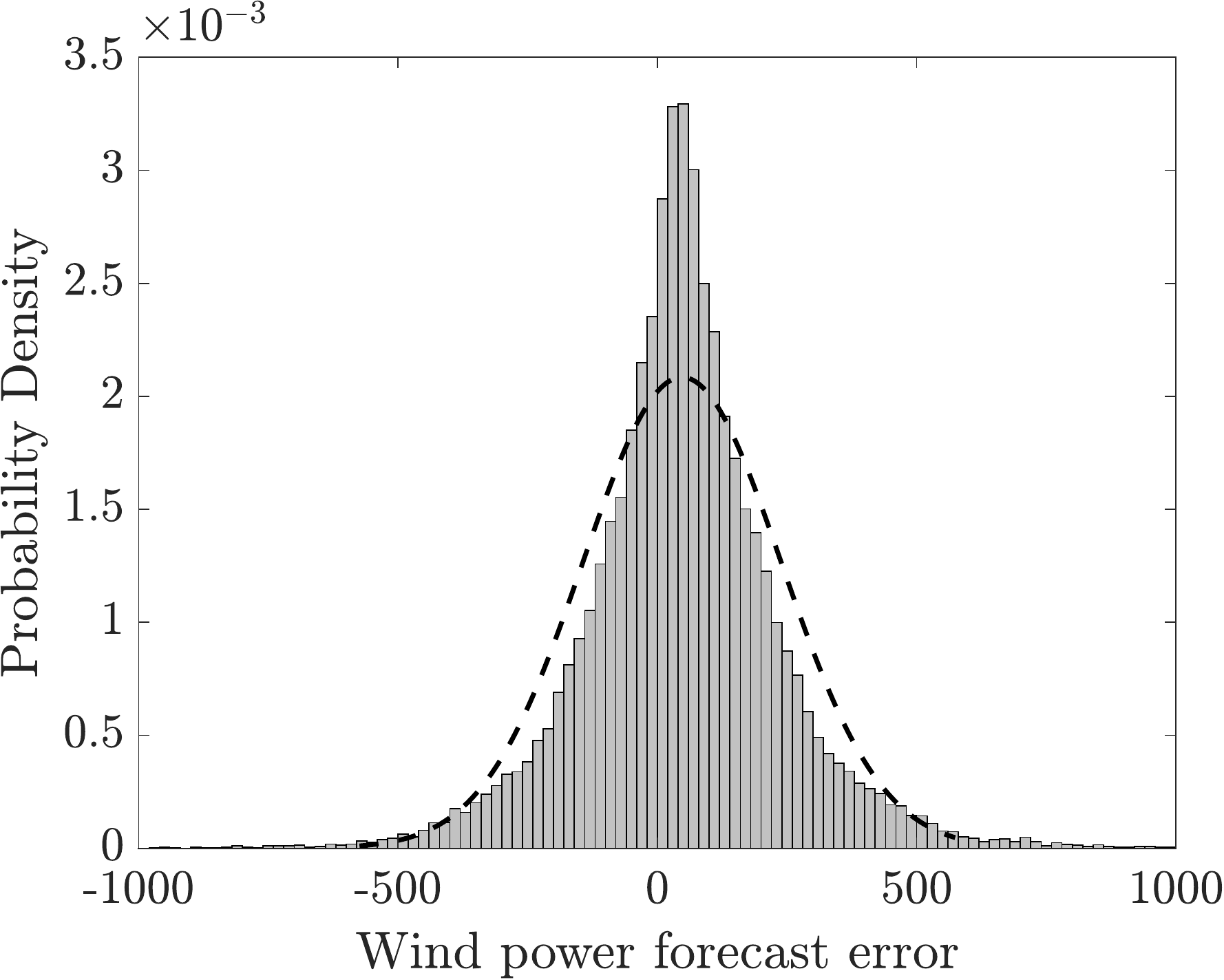}
    \caption{Distribution of forecasting error in wind power of aggregate Belgian wind farms vs a normal distribution.}
  \label{fig:Errors}
\end{figure}

Remark~\ref{rem:dist} is relevant in the case of wind generation, since as discussed in~\citet{lange2005uncertainty, hodge2011wind}, wind speed forecast errors closely follow  a normal distribution. However, due to the non-linear relationship between wind speed and wind power generated, the errors in wind power forecast do not closely follow a normal distribution. In particular, the distribution has a larger peak at zero, and fatter tails than a normal distribution. However, as discussed in~\citet[][Section III.D]{hodge2011wind}, the wind power forecast error distributions are close to being symmetric. To illustrate these points, below we provide the distribution generated by the day-ahead forecasting errors in the aggregated Belgian wind farms (\url{http://www.elia.be/en/grid-data/power-generation/wind-power}) between  01/11/2018 and 01/11/2019, that have an skewness of $-0.11$. These facts are illustrated in Figure~\ref{fig:Errors}, which matches similar plots in~\citet{lange2005uncertainty, hodge2011wind}.

Another assumption made about the distribution of $\Delta \W_n^f$, $n \in \B$ is that $\mathbb{E}(\Delta \W_n^f)=0$. However, if $\mathbb{E}(\Delta \W_n^f)$ is known, one can set $W_n^f \to W_n^f + \mathbb{E}(\Delta\W_n^f)$, and $\Delta\W_n^f \to \Delta\W_n^f - \mathbb{E}(\Delta\W_n^f)$.

Throughout, we have assumed that $0 < \epsilon < 0.5$ (cf., discussion following~\eqref{eq:CCmodel}
). This ensures that the $1-\epsilon$ quantile of a continuous symmetric distribution is greater than zero. This condition is needed for the prices in~\eqref{eq:tau} to be well defined. However, $\epsilon$ stands for the tolerance for constraint violations, so in practice, it will always hold that $\epsilon$ is definitely lower than $0.05$. Another necessary condition for the prices in~\eqref{eq:tau} to be well defined is that $\sigma_n > 0$ for all $n \in \B$, which means that all buses should have an uncertain source of power. However, this assumption can be dropped by making a small change to the DCCO market-clearing model~\eqref{eq:detCCmodel} and the prices defined in~\eqref{eq:tau}.

\begin{remark}
\label{rem:zero}
Let $n' \in \B$ be a node in which the standard deviation of $\Delta \W_n^f$ is zero; that is, $\sigma_{n'} = 0$. Then, the DCCO model~\eqref{eq:detCCmodel} and Definition~\ref{def:CCpricing} are correct, and Theorem~\ref{thm:CCpricing} holds, after making the following changes: the constraints $\alpha_i^u = 0, \alpha_i^d = 0$ should be added to~\eqref{eq:detCCmodel}, and in~\eqref{eq:tau}, the prices should be changed to $\tau_i^u = {y_i^u}^*$, $\tau_i^d = {y_i^d}^*$, for all $i \in \I_{n'}$.  
\end{remark}

This remark follows after noticing that in~\eqref{eq:tauu}, if ${\alpha_i^u}^* = 0$, we need to show that $\tau_i^u{r_i^u}^* \ge {y_i^u}^*{r_i^u}^*$, which clearly follows after setting $\tau_i^u = {y_i^u}^*$. A similar argument can be used to justify the changes related to $\alpha_i^d$ and $\tau_i^d$. Note that in Sections~\ref{sec:casestudy} and~\ref{sec:numerical}, we consider an electricity market in which there is no VRES generation in one of the buses.

The most important assumption we make in Section~\ref{sec:CCmodel} is that the distribution of $\Delta \W_n^f$ is symmetric for all~$n \in \B$ (an assumption also made in~\citet{pozo2012chance, bienstock2014chance, lubin2015robust}). In this case, for any $a, b \in \R$, enforcing the probabilistic constraint $\mathbb{P}(a\Delta \W_n^f \le b) \ge 1-\epsilon$ is equivalent to enforcing that $b- a\sigma_n \mathbb{Q}(\widetilde{\Delta \W}_n^f)_{1-\epsilon} \ge 0$, regardless of the sign of $a$, thanks to the fact that  the $\Delta \W_n^f$ distribution's symmetry implies that $\mathbb{Q}(\widetilde{\Delta \W}_n^f)_{1-\epsilon} = -\mathbb{Q}(\widetilde{\Delta \W}_n^f)_{\epsilon}$. Thus, the affine controls in~\eqref{eq:controls} can be assumed to be non-negative without loss of generality. On the other hand, if the distribution of $\Delta \W_{n'}^f$ is not symmetric for some $n' \in \B$, then the  deterministic reformulation of~\eqref{eq:CCmodel} would depend on the signs of the affine controls used by the market participants located in bus $n'$ (e.g., $\alpha_i^u$ for all $i \in \I_{n'}$). This could be handled by adding binary variables to~\eqref{eq:detCCmodel}. Finding the corresponding revenue adequate pricing scheme for this type of market is an interesting line of research that relates to results regarding the handling of market non-convexities~\citep[see, e.g.,][]{ref.oneill2005, liberopoulos2016critical, ref.ruiz2012pricing}.

\section{Summary of SO Market-Clearing model and Pricing Scheme}
\label{sec:stocmodel}
Following~\citet{morales2012pricing}, assume that the uncertainty in VRES power generation at each bus~$n \in \B$ of the network is given by:
\begin{equation}
\label{eq:stocwinddef}
\W_n \sim \{(W_n^{\omega}, \pi^{\omega}), \forall \omega \in \Omega\},
\end{equation}
where $\Omega$ is a finite scenario set, $W_n^{\omega}$ is the VRES power generated at bus $n \in \B$ in scenario~$\omega$, occurring with probability $\pi^{\omega} > 0$ for all $\omega \in \Omega$, with $\sum_{\omega \in \Omega} \pi^{\omega} = 1$. Also, consider the additional relevant notation given in Table~\ref{tab:stoch}.

\begin{table}[!htb]
\caption{Additional nomenclature for the stochastic optimization market-clearing model.}
\label{tab:stoch}
\begin{center}
\begin{tabular}{lp{0.85\linewidth}}
		     \toprule
\multicolumn{2}{c}{\bf Variables}\\
\midrule	
		     $\delta_{n}^{\omega}$ & Real-time voltage angle at bus $n \in \B$, in scenario $\omega \in \Omega$\\	
			 $r_i^{u\omega}$ & Real-time upward reserve deployed by generator $i \in \I$, in scenario $\omega \in \Omega$\\
			 $r_i^{d\omega}$ &Real-time downward reserve deployed by generator $i \in \I$, in scenario $\omega \in \Omega$\\
			 $s_j^{\omega}$ & Real-time involuntary curtailment of scheduled load $j \in \J$, in scenario $\omega \in \Omega$ \\
			 $w_n^{\spill,\omega}$ & Real-time VRES power spilled at bus $n \in \B$, in scenario $\omega \in \Omega$\\
\bottomrule
\end{tabular}
\end{center}
\end{table}

 Using~\eqref{eq:stocwinddef}, the SO model associated with the deterministic market-clearing model~\eqref{eq:detmodel} is given by:
	\begin{subequations}\label{eq:stocmodel}
		\begin{align}
\min & \sum_{\omega \in \Omega} \pi^{\omega} \left (\sum_{i \in \I} \Bigg (C_i p_i + C_i^u r_i^{u\omega} - C_i^d r_i^{d\omega} \right )+ \nonumber\\ 
& \hspace{2in}\sum_{n \in \B} C_n^w (W_n^{\omega}- w_n^{\spill,\omega}) + \sum_{j \in \J} V_js_j^{\omega} \Bigg)\label{eq:stoc_obj} \\
		\st   &\sum_{i \in \I_n} p_i + w_n^{\sched} - \sum_{(n,l) \in \L} B_{nl} (\delta_{n}^0 -  \delta_{l}^0) =\sum_{j \in \J_n}L_j , \label{eq:stoc_balance} & (\lambda_n) && \forall n\in \B\\
		& \sum_{i \in \I_n}(r_i^{u\omega} - r_i^{d\omega}) + \sum_{j \in \J_n} s_j^{\omega} + (W_n^{\omega} - w_n^{\sched} - w_n^{\spill,\omega})  + \nonumber\\
		& \hspace{2in}\sum_{(n,l) \in \L} B_{nl} (\delta_{n}^0 - \delta_{n}^{\omega} - \delta_{l}^0 + \delta_{l}^{\omega}) = 0,   \label{eq:stoc_reblanace}& (\nu_n^{\omega}) && \forall n\in \B, \omega \in \Omega\\
		& B_{kl}(\delta^0_{k} - \delta^0_{l}) \leq   \overline{C}_{kl}, B_{kl}(\delta_{k}^{\omega}-\delta_{l}^{\omega}) \leq   \overline{C}_{kl}, \label{eq:stoc_line2} &&& \hspace{-2em}\forall (k,l) \in \L, \omega \in \Omega\\
		& -w_n^{\spill,\omega} \ge -W_n^{\omega}, \label{eq:stoc_spillbd} &&& \forall n \in \B, \omega \in \Omega\\
		&  -w_n^{\sched} \geq  -\overline{W}_n,~-p_i  \geq -\overline{P}_i, \label{eq:stoc_wind_limit} &  && \forall n \in \B, i \in \I\\ 
		& -r_i^{u\omega} \geq -\overline{R}_i^u,~-r_i^{d\omega} \geq  -\overline{R}_i^d,\label{eq:stoc_rampup_limit} &&&\forall i \in \I, \omega \in \Omega    \\
& p_i + r_i^{u\omega} - r_i^{d\omega} \geq 0, -p_i -  r_i^{u\omega} + r_i^{d\omega} \geq -\overline{P}_i, \label{eq:stoc_con1_limit} &&& \forall i \in \I, \omega \in \Omega     \\
		& -s_j^{\omega} \geq -L_j. \label{eq:stoc_demand_limit} &&& \forall j \in \J, \omega \in \Omega   \\
				&p_i,  r_i^{u\omega}, r_i^{d\omega} \ge 0, \label{eq:stoc_nonneg1}&&& \forall i \in \I,\omega \in \Omega\\
		&w_n^{\sched}, w_n^{\spill, \omega} \geq 0, \label{eq:stoc_nonneg2}&&& \forall n \in \B, \omega \in \Omega	 \\
		&s_j^{\omega} \geq 0. \label{eq:stoc_nonneg3}&&& \forall j \in \J, \omega \in \Omega
		\end{align}
	\end{subequations}
Similar to the DCCO market-clearing model~\eqref{eq:detCCmodel}, the SO market-clearing model~\eqref{eq:stocmodel} is a LP, in which the dual variables associated with
the bus power balance constraints in the scheduling~\eqref{eq:stoc_balance} and real-time~\eqref{eq:stoc_reblanace} stages are labeled in parenthesis at the right of the constraints (where for comparison purposes, we abuse notation by using the same labels used in~\eqref{eq:detCCmodel}). As shown by~\citet{morales2012pricing}, the optimal values of these dual variables can be used to define a revenue adequate SO pricing scheme for the SO market-clearing model~\eqref{eq:stocmodel}. Analogous to Section~\ref{sec:CCPricing}, in what follows, for any primal or dual variable $(\cdot)$ of the SO market-clearing model~\eqref{eq:stocmodel},~$(\cdot)^*$ indicates the optimal value of the decision variable.

\begin{definition}[{Stochastic Optimization Pricing Scheme~\citep{morales2012pricing}}] 
\label{def:stocpricing} Energy transactions settled via the SO market-clearing model~\eqref{eq:stocmodel}
are priced as follows:
\begin{enumerate}[label =(\roman*)]
	\item Each conventional generator $i \in \I_n, n\in \B$, is compensated at price
	$\lambda_{n}^*$ 
	for every unit of scheduled power ${p_i}^*$. VRES generators at bus $n \in \B$ are compensated at price $\lambda_n^*$ for every unit of scheduled power ${w_n^{\sched}}^*$.
	\label{it:stocpricegens}
	\item Each load $j \in \J_n, n \in \B$ is charged at price $\lambda_n^*$ for every unit of its scheduled consumption ${L_j}$.
	\item Each conventional generator $i \in \I_n, n\in \B$ is compensated (resp., charged) 
	at price $\frac{{\nu_n^{\omega}}^*}{\pi^{\omega}}$ in scenario $\omega \in \Omega$
	for every unit of real-time upward reserve deployed  ${r^{u\omega}_i}^*$ (resp., downward reserve deployed ${r^{d\omega}_i}^*$).
	\label{it:stocpriceloads}
	\item VRES generators at bus $n\in \B$ are compensated (resp., charged) at price $\frac{{\nu_n^{\omega}}^*}{\pi^{\omega}}$ in scenario $\omega \in \Omega$ for every unit of surplus (resp., shortage) of real-time net VRES power generated over (resp., under) the VRES power generation scheduled $\max\{0,W_n^{\omega} - {w_n^{\sched}}^*- {w_n^{\spill, \omega}}^* \}$ (resp., $\max\{0,-W_n^{\omega} + {w_n^{\sched}}^*+ {w_n^{\spill, \omega}}^*\}$).
	\label{it:stocpricewinds}
	\item Each load $j\in\J_n, n \in \B$ is compensated at price $\frac{{\nu_n^{\omega}}^*}{\pi^{\omega}}$ in scenario $\omega \in \Omega$ for every unit of real-time involuntary load curtailment ${s_j^{\omega}}^*$.
	\label{it:stocpricesheds}
\end{enumerate}
\end{definition}

As shown in~\citet{morales2012pricing}, the stochastic pricing scheme in Definition~\ref{def:stocpricing} guarantees that revenue adequacy (or cost recovery) in expectation is satisfied for the power generators and the market administrator in the SO market-clearing model~\eqref{eq:stocmodel}.

\begin{theorem}[\citet{morales2012pricing}]
\label{thm:SOpricing}
The SO pricing scheme introduced in Definition~\ref{def:stocpricing} guarantees that in the SO market-clearing model~\eqref{eq:stocmodel}, the market administrator is ensured revenue adequacy in expectation (i.e., has non-negative profit in expectation), and the generating units in the market, including the VRES generators, are ensured cost recovery in expectation (i.e., their revenue is greater than or equal than their operating costs in expectation).
\end{theorem}

 \section{Expected profit and standard deviation for stochastic pricing scheme.}
 \label{sec:stocrevs}
We next present the expected value and the standard deviation of the profit of all the market participants in the SO pricing scheme, for the purpose of illustratively discussing the revenue adequacy property in the CCO pricing scheme introduced in Definition~\ref{def:CCpricing} and the SO pricing scheme in Definition~\ref{def:stocpricing}. These results readily follow from~\citet[][Appendix]{morales2012pricing}.

For any $n \in \B, \omega \in \Omega$, let 
\[
A^{\omega}_n = \dsum_{i \in \I_n} ({r_i^{u\omega}}^* - {r_i^{d\omega}}^*) - ({w_n^{\spill,\omega}}^*+ {w_n^{\sched}}^* - W_n^{\omega})+ \dsum_{j \in \J_n} {s_j^{\omega}}^*.
\]
Then, the expected profit of the market administrator in the SO market-clearing pricing scheme is given by
\begin{equation}
\label{eq:stocISOexprevenue}
\hat{\Gamma}^o = 
-\dsum_{n \in  \B} \lambda_n^* \left (\dsum_{i \in \I_n} p_i^* + {w_n^{\sched}}^* - \dsum_{j \in \J_n} L_j \right )
- \dsum_{n \in  \B} \dsum_{\omega \in \Omega} {\nu_n^{\omega}}^*A^{\omega}_n,
\end{equation}
and the standard deviation of the profit for the market administrator is given by
\begin{equation}
\label{eq:stocISOstd}
\hat{\sigma}^o = \sqrt{\dsum_{\omega \in \Omega} \pi^{\omega} \left ( \dsum_{n \in  \B}  \frac{{\nu_n^{\omega}}^*}{\pi^{\omega}}A_n^{\omega} \right )^2 - \left (\dsum_{\omega \in \Omega} \pi^{\omega} \left ( \dsum_{n \in  \B}  \frac{{\nu_n^{\omega}}^*}{\pi^{\omega}}A_n^{\omega} \right ) \right )^2}.
\end{equation}

For any $n \in \B, \omega \in \Omega$, let
\[
B_n^{\omega} = \frac{{\nu_n^{\omega}}^*}{\pi^{\omega}}(W_n^{\omega} -  {w_n^{\sched}}^*  -  {w_n^{\spill,\omega}}^*) - C_n^w(W_n^{\omega} -  {w_n^{\spill,\omega}}^*).\]
Then, for all $n \in \B$, the expected profit of the VRES generators at bus $n$ in the SO market-clearing pricing scheme is given by
\begin{equation}
\label{eq:stocWINDexprevenue}
\hat{\Gamma}^{w_n} =  \lambda_{n}^*{w_n^{\sched}}^* +  \sum_{\omega \in \Omega} \pi^{\omega}B^{\omega}_n,
\end{equation}
and the standard deviation of the VRES generator's profit at bus $n \in \B$ is given by
\begin{equation}
\label{eq:stocWINDstd}
\hat{\sigma}^{w_n} = \sqrt{\sum_{\omega \in \Omega} \pi^{\omega}(B^{\omega}_n)^2 - \left (\sum_{\omega \in \Omega} \pi^{\omega} B^{\omega}_n \right)^2}.
\end{equation}

For any $i \in I_n$, $n \in \B$, $\omega \in \Omega$, let
\[
D^{\omega}_{in} = \left(\frac{{\nu_n^{\omega}}^*}{\pi^{\omega}} - C_i^u \right ){r_i^{u\omega}}^*
-  \left(\frac{{\nu_n^{\omega}}^*}{\pi^{\omega}} - C_i^d \right ){r_i^{d\omega}}^*.
\]
Then, for all $i \in \I_n$, $n \in \B$, the expected profit of conventional generator $i$th in the 
SO market-clearing pricing scheme is given by
\begin{equation}
\label{eq:stocGENexprevenue}
\hat{\Gamma}^{g_i} = (\lambda_n^* - C_i)p_i^* + \sum_{\omega \in \Omega} \pi^{\omega} D^{\omega}_{in},
\end{equation}
and the standard deviation of the profit for the $i$th conventional generator is given by
\begin{equation}
\label{eq:stocGENstd}
\hat{\sigma}^{g_i} = \sqrt{ \sum_{\omega \in \Omega} \pi^{\omega} 
\left ( D^{\omega}_{in} \right)^2- \left (\sum_{\omega \in \Omega} \pi^{\omega} 
 D^{\omega}_{in} \right)^2}.
\end{equation}

For all $j \in \J_n$, $n \in \B$, the expected surplus of consumer $j$th in the SO market-clearing pricing scheme is given by
\begin{equation}
\label{eq:stocCUSTexprevenue}
\hat{\Gamma}^{L_j} =  \sum_{\omega \in \Omega} {\nu_n^{\omega}}^*{s_j^{\omega}}^*- \lambda_n^*L_j,
\end{equation}
and the standard deviation of the consumer surplus for the $j$th load is given by
\begin{equation}
\label{eq:stocCUSTstd}
\sigma^{L_j} = \sqrt{\sum_{\omega \in \Omega} \pi^{\omega}\left (\left(\frac{\nu_n^*}{\pi^{\omega}} \right){s_j^{\omega}}^*\right )^2 - \left (\sum_{\omega \in \Omega} \pi^{\omega}\left (\frac{\nu_n^*}{\pi^{\omega}} \right){s_j^{\omega}}^*\right )^2}.
\end{equation}

\end{appendices}
\end{document}